\documentclass[10pt,journal]{IEEEtran}
\usepackage{amsmath,amssymb,amsfonts}
\usepackage{graphicx}
\usepackage{multirow}
\usepackage{booktabs}
\usepackage{cuted}
\usepackage{CJK}
\usepackage{indentfirst}
\usepackage{xcolor}
\usepackage{hyperref}
\usepackage{epstopdf}
\usepackage{bm}
\usepackage{cases}
\usepackage{subfig}
\usepackage{caption}
\usepackage{bookmark}
\usepackage{makecell}

\usepackage{algorithmic}
\usepackage{algorithm}
\usepackage{array}
\usepackage{textcomp}
\usepackage{stfloats}
\usepackage{url}
\usepackage{verbatim}
\usepackage{cite}
\hypersetup{
colorlinks=true,
linkcolor=blue,
citecolor=blue
}
 \usepackage{hyperref}

\newtheorem{remark}{Remark}

\begin{document}

\title{Waveform Bandwidth and Frequency Diversity in FDA-MIMO Range--Angle Estimation: A CRLB Perspective}

\author{Guoqing Xia, Qu Luo, \textit{Member, IEEE}, Yu Liu, Pei Xiao, \textit{Senior Member, IEEE}, Cicek Cavdar, Member, IEEE, Emil Björnson, Fellow, IEEE.}

\maketitle


\begin{abstract}
Frequency-diverse array multiple-input multiple-output (FDA-MIMO) radar
introduces range-dependent phase diversity through element-dependent
carrier-frequency offsets, in addition to the delay information carried
by transmitted waveforms. This paper develops a bandwidth-aware
Cram\'er--Rao lower bound (CRLB) framework to characterize how these
two mechanisms jointly affect range--angle estimation. Starting from
a Nyquist-rate sampled observation model that retains waveform delays
and FDA frequency offsets, we derive an equivalent Fisher information
matrix (EFIM) after eliminating an unknown quasi-static complex
scattering coefficient as a nuisance parameter. We derive
closed-form CRLB approximations that reveal the interplay between
waveform root-mean-square (RMS) bandwidth and FDA frequency diversity.
A three-factor decomposition of the squared normalized
cross-information further characterizes range--angle coupling.
Results show that when increased independently, both waveform RMS bandwidth and FDA
frequency diversity reduce the range CRLB, but FDA frequency diversity increases the
angle CRLB at fixed nonzero waveform bandwidth, whereas
waveform RMS bandwidth reduces it. Additionally, under a fixed aggregate spectral
budget, the range CRLB exhibits an interior maximum, whereas the angle CRLB increases
monotonically with the FDA spectral-allocation fraction. Numerical
results corroborate the closed-form approximations and the predicted
trends.
\end{abstract}



\section{Introduction}
\label{sec:intro}

Joint range--angle estimation is a fundamental task in radar sensing.
In conventional array radar, angular information is provided by
aperture-dependent spatial responses, whereas range information arises
from the propagation-delay sensitivity of the transmitted waveform,
which is governed by its root-mean-square (RMS) bandwidth
\cite{vanTrees2002OptimumArrayProcessing,Dogandzic2001TSPActiveArray}.
Colocated multiple-input multiple-output (MIMO) radar further exploits
distinct transmit waveforms to synthesize a virtual aperture and
combine spatial and waveform diversity
\cite{Li2007,Li2008TSPRangeCompression}.
These mechanisms form the basis for examining the additional range
information introduced by frequency-diverse arrays.

A frequency-diverse array (FDA) assigns slightly different carrier
frequencies to its transmit elements, thereby introducing
range-dependent inter-element phase differences in addition to the
conventional angular dependence
\cite{Antonik2006,Secmen2007,Wang2015APM,Wang2016IET}.
The resulting range--angle-dependent responses and their
time-varying characteristics have been extensively studied,
including corrections to previously reported beampattern models
\cite{Zubair2020TSP,Chen2019TAPTimeInvariant,Tan2021TAPTimeCorrection}.
Frequency-offset arrangements have also been developed to address
range--angle coupling and ambiguity in FDA responses
\cite{Liu2017JSTSP,Wang2018TAESAmbiguity}.
However, the range--angle dependence of FDA array responses
does not directly quantify the achievable estimation accuracy.
Joint range--angle estimation performance also depends on the
information associated with each parameter and their
cross-information, motivating a performance bound analysis.

FDA-MIMO radar combines element-dependent carrier frequencies with
the waveform diversity of colocated MIMO radar
\cite{Sammartino2013TAES}.
Xu \emph{et al.}~\cite{Xu2015TSP} developed a joint range--angle
estimation framework and derived the corresponding Cram\'{e}r--Rao
bounds (CRLBs), including the effects of parameter coupling.
Xiong \emph{et al.}~\cite{Xiong2018TAES} further investigated
CRLB, mean-square-error, and resolution characteristics.
Subsequent studies have addressed single-snapshot estimation and
alternative FDA array configurations
\cite{Lan2021TAESSS,Mao2022RS,Wang2025RS}.
More recently, Sun \emph{et al.}~\cite{Sun2024TAESOffset}
analyzed the effects of transmit- and receive-side frequency-offset
errors on range--angle estimation and derived the corresponding CRLBs.
For moving targets, Wang \emph{et al.}~\cite{Wang2025TVTZZB}
examined multidimensional parameter estimation and Ziv--Zakai bounds,
accounting for coupling between bandwidth-level frequency offsets
and target angle and velocity.
The CRLB has also served as a design criterion for FDA subarray and
hybrid transmitter architectures
\cite{Wang2015LAWPSubarrayCRLB,Gao2016IETFDAHybrid}, as well as
frequency-offset optimization
\cite{Cheng2023TAESOptimalOffsets,Lan2025SPLFreqIncOpt}.
These studies demonstrate the influence of frequency diversity,
array architecture, and frequency-offset imperfections on
joint estimation performance.

From a signal-model perspective, FDA-MIMO ranging involves two
distinct information mechanisms. The first is the delay sensitivity
of the waveform envelope, governed by its RMS bandwidth
\cite{Dogandzic2001TSPActiveArray,Li2008TSPRangeCompression}.
The second is the range-dependent carrier-phase variation induced by
transmit frequency offsets \cite{Xu2015TSP,Xiong2018TAES}.
Although both mechanisms contribute to range sensitivity, their
effects on the resulting CRLBs are not simply additive. In particular,
when range and angle are jointly estimated in the presence of an
unknown complex scattering coefficient, the effective range
information also depends on its coupling with angle. Thus, an
increase in range sensitivity alone does not necessarily imply a
commensurate improvement in the marginal estimation bound.

Wang \emph{et al.}~\cite{Wang2014CRLB} analyzed CRLBs
based on the received signal before and after matched filtering and explicitly
examined the role of waveform bandwidth. Geng \emph{et al.}~
\cite{Geng2023IETMulticarrier} investigated multicarrier FDA-MIMO
transmission, including range--angle CRLB analysis and frequency-component
design for beampattern synthesis. Liu \emph{et al.}~\cite{Liu2025TAESEPC}
further investigated element-pulse-coded FDA-MIMO radar, employing
frequency-domain subband synthesis to mitigate range ambiguity and
enhance range resolution, together with range--angle CRLB analysis.
These contributions establish the relevance of waveform and
frequency-offset parameters, although their signal representations
and design objectives differ. A complementary question is how
waveform-delay and FDA-induced phase sensitivities jointly determine
the equivalent Fisher information matrix (EFIM) when the unknown
scattering coefficient is eliminated, and how this interaction can be
expressed in a tractable form. Addressing this question requires a
model that retains both dependencies explicitly, without presupposing
a reduced-dimensional steering-vector or matched-filter-output
representation.

A related issue arises when waveform bandwidth and FDA frequency
diversity share a finite spectral resource. In such case, these two parameters
are interdependent, and their individual effects on
estimation accuracy cannot be directly extended to the jointly
constrained setting. This motivates a unified characterization of the two
information mechanisms and their tradeoffs under an aggregate
spectral constraint.

In this paper, we develop a bandwidth-aware CRLB framework for joint
range--angle estimation in a colocated monostatic FDA-MIMO radar.
Starting from the delayed received waveforms, we derive a sampled
observation model that explicitly retains waveform-delay and
FDA-induced phase dependencies. We then characterize the EFIM for an
unknown quasi-static complex scattering coefficient, derive
closed-form CRLB approximations, and examine their dependence on
system parameters and finite spectral resources. The main
contributions are summarized as follows.

\begin{itemize}
\item \emph{Waveform-resolved EFIM:}
We formulate a Nyquist-rate sampled FDA-MIMO model that retains
waveform delays, carrier-frequency offsets, and array geometry.
Eliminating the unknown complex scattering coefficient through a
Schur complement yields the EFIM for joint range--angle estimation.

\item \emph{Closed-form CRLBs and information interplay:}
Under specified waveform-orthogonality, spectral-centering, and
centered-array conditions, we derive closed-form approximations
to the range and angle CRLBs that explicitly characterize
the interplay between waveform RMS bandwidth and FDA frequency
diversity in estimation accuracy. A three-factor decomposition
of the normalized cross-information further reveals their
effects on range--angle coupling.

\item \emph{Finite-spectrum tradeoffs:}
Under a centered linear frequency-offset law and a waveform-family
occupied-bandwidth approximation, we characterize CRLB trends subject
to a fixed aggregate spectral budget. The range CRLB exhibits an
interior maximum and a boundary-optimal allocation, whereas the angle
CRLB increases with the FDA spectral share. Numerical results
validate the analytical trends.
\end{itemize}

The remainder of this paper is organized as follows.
Section~\ref{sec:model} presents the FDA-MIMO signal model under
quasi-static radar cross section (RCS). Section~\ref{sec:modelII}
derives the unified CRLB and the Schur-complement reduction.
Section~\ref{sec:CF CRLB} establishes the closed-form CRLB
representation. Section~\ref{sec:CRLB_analysis} analyzes the resulting
parameter dependence and spectral-resource tradeoffs.
Section~\ref{sec:numerical_validation} presents numerical validation,
and Section~\ref{sec:conclusion} concludes the paper.
A conditional-average FIM interpretation for sample-varying RCS is
provided in Appendix~\ref{sec:fast_extension}.

\emph{Notation}: $\mathbb{R}$ and $\mathbb{C}$ denote the fields of
real and complex numbers, respectively. Scalars are denoted by
lower-case letters, and vectors and matrices by lower- and upper-case
boldface letters, respectively. ${\bm I}$ denotes an identity matrix. $\mathfrak{R}\{\cdot\}$ and
$\mathfrak{T}\{\cdot\}$ denote the real and imaginary parts,
respectively. $\mathcal{CN}(\cdot,\cdot)$ denotes a proper complex Gaussian distribution. The conjugate, transpose, and conjugate transpose are
denoted by $(\cdot)^*$, $(\cdot)^{\rm T}$, and $(\cdot)^{\rm H}$,
respectively. $\mathbb{E}\{\cdot\}$ denotes mathematical
expectation. $|\cdot|$ takes absolute value or modulus. ${\rm diag}\{\cdot\}$
forms a diagonal matrix from its arguments. $\|\cdot\|_2$ denotes
the $\ell_2$ norm. The notation $a\ll b$ means that $a$ is much smaller
than $b$.


\section{System Model}
\label{sec:model}

We consider a colocated monostatic FDA-MIMO radar observing a single
stationary far-field point target within one coherent processing interval.
The received-signal model is constructed from the transmitted complex
analytic signals and their propagation delays, retaining the waveform
delays, element-dependent frequency offsets, and transmit--receive array
geometry explicitly.

\subsection{Transmit Signal}

\begin{figure}[!t]
    \centering
    \includegraphics[width=\linewidth]{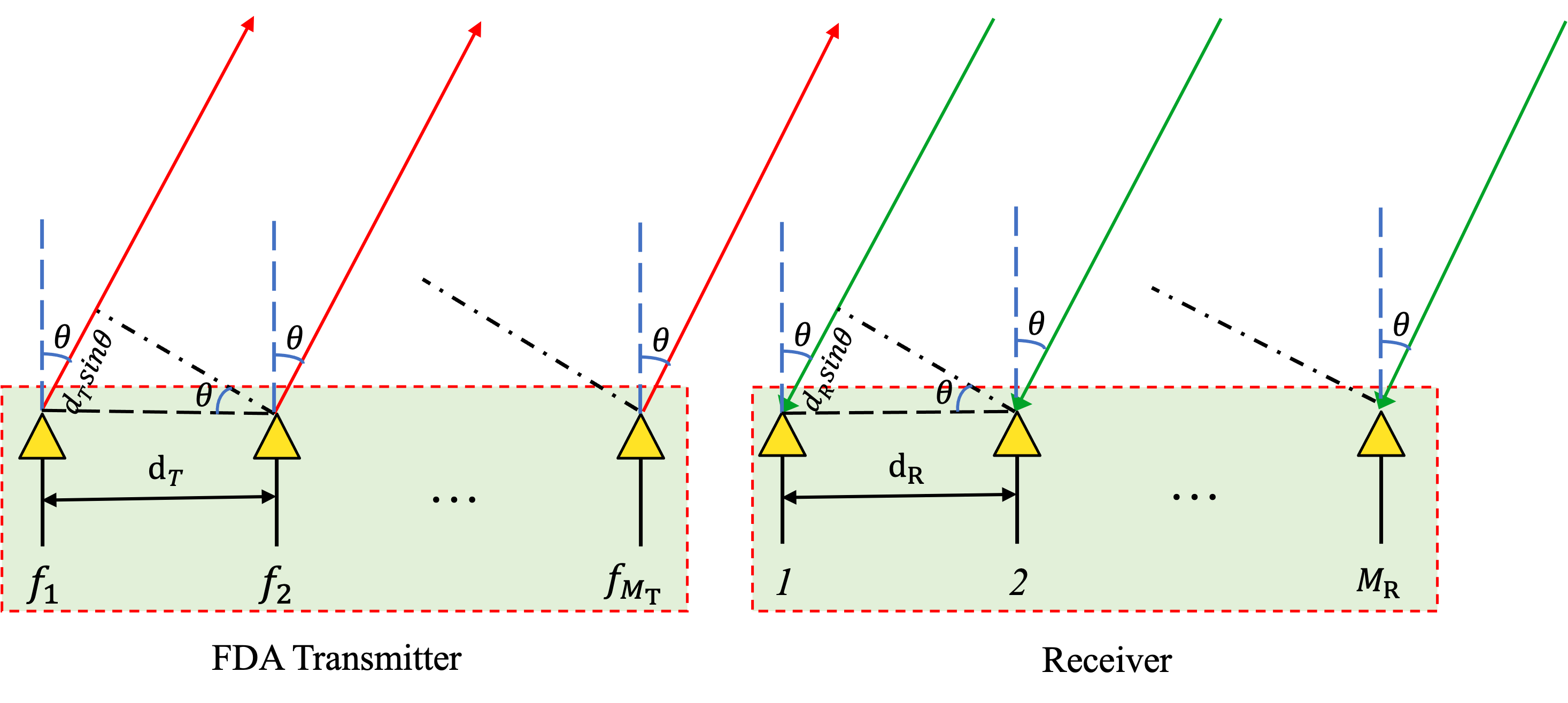}
    \caption{Colocated monostatic FDA-MIMO geometry under the far-field
approximation.}
    \label{fig:fda_mimo}
\end{figure}

As shown in Fig.~\ref{fig:fda_mimo}, the transmit and receive uniform
linear arrays (ULAs) are aligned along the same axis, with
$M_{\rm T}$ transmit elements and $M_{\rm R}$ receive elements,
respectively. Their element spacings are $d_{\rm T}$ and $d_{\rm R}$.

The $m$-th transmit element radiates
\begin{equation}
x_m(t) = \psi_m(t)\, e^{j2\pi f_m t},
\end{equation}
where $\psi_m(t)$ denotes the complex baseband waveform
transmitted by the $m$th element, and
\begin{equation}
f_m=f_c+\Delta f_m.
\end{equation}
Here, $f_c$ is the reference carrier frequency and $\Delta f_m$ is the element-dependent frequency offset. The frequency offsets $\{\Delta f_m\}_{m=1}^{M_{\rm T}}$ introduce element-dependent carrier frequency diversity. 

Let $\mathcal{B}_m$ denote the nominal occupied baseband spectral
support of the $m$th waveform. Frequency shifting moves this support
to $\mathcal{B}_m+\Delta f_m$, so the aggregate spectral support is
\begin{equation}
\mathcal{B}_{\rm agg}
\triangleq
\bigcup_{m=1}^{M_{\rm T}}
\left(
\mathcal{B}_{m}+\Delta f_m
\right).
\end{equation}
We denote by $B_{\rm agg}$ the width of the smallest contiguous
frequency interval containing $\mathcal{B}_{\rm agg}$, i.e.,
$B_{\rm agg}=\sup\mathcal{B}_{\rm agg}-\inf\mathcal{B}_{\rm agg}$.

Following the standard FDA narrow-offset condition
\cite{Wang2014CRLB,Chen2019TAPTimeInvariant,
Tan2021TAPTimeCorrection}, we assume
\begin{equation}
|\Delta f_m| \ll f_c,\qquad m=1,\ldots,M_{\rm T}.\label{eq:small_deltaf}
\end{equation}

At the transmitter level, the representation includes several
standard array-radar architectures as special cases:
\begin{itemize}
\item If $\Delta f_m=0$ for all $m$, the model reduces to a
colocated MIMO radar \cite{Li2007}.
\item If $\Delta f_m=0$ and
$\psi_m(t)=w_m\psi(t)$, it reduces to a digital phased-array radar
\cite{talisa2016benefits}.
\item If $\psi_m(t)=\psi(t)$ and $\Delta f_m\neq 0$, it describes a
coherent FDA transmitter \cite{Gui2018TSPCoherentPulsedFDA}.
\end{itemize}
These transmitter special cases do not necessarily satisfy the waveform
cross-orthogonality conditions introduced in Section~\ref{sec:CF CRLB}.

Define the transmitted signal vector as
\begin{equation}
\mathbf{x}(t) = \mathbf{W}(t)\boldsymbol{\psi}(t)e^{j2\pi f_c t},\label{eq:transmit vector}
\end{equation}
where $\boldsymbol{\psi}(t)$ is the baseband waveform vector and
$\mathbf W(t)$ is the diagonal matrix of frequency-offset-induced
phases, given by
\begin{equation}
\mathbf{W}(t) = \mathrm{diag}\big(e^{j2\pi \Delta f_1 t}, \ldots, e^{j2\pi \Delta f_{{M_{\rm T}}} t}\big).
\end{equation}
Equation~\eqref{eq:transmit vector} represents FDA transmission as
antenna-dependent frequency modulation of the baseband waveform vector.
Such frequency diversity can be combined with multicarrier modulation
or space--time coding
\cite{Zhang2025AFDM,Wachowiak2025JCAS,Wang2021TSPSpaceTimeCoding}. 

\subsection{Received Signal}

Consider a far-field target at angle $\theta$, measured from the
array broadside. Denote $T_c$ and
$R_c$ as the reference points of the transmit and receive arrays,
respectively. They are introduced as coordinate references and are not
required to coincide with physical antenna locations. The range
parameter $r$ is defined through the reference transmit--target--receive
path associated with $(T_c,R_c)$, namely
\begin{equation}
2r=r_{{\rm T},c}+r_{{\rm R},c},
\end{equation}
where $r_{{\rm T},c}$ and $r_{{\rm R},c}$ are the propagation distances from $T_c$ to the target and from the target to $R_c$, respectively. 

The transmit and receive elements are indexed relative to their
array centers as
\begin{equation}
i_{{\rm T},m}=m-\frac{M_{\rm T}+1}{2},\qquad m=1,\ldots,M_{\rm T},\label{eq:centered_transmit_index}
\end{equation}
and
\begin{equation}
i_{{\rm R},n}=n-\frac{M_{\rm R}+1}{2},\qquad n=1,\ldots,M_{\rm R}.\label{eq:centered_receive_index}
\end{equation}
The array-center reference coordinate is zero for both arrays.
For an odd array size, the zero index corresponds to a physical center
element; for an even array size, it specifies the virtual phase center
and does not correspond to an indexed antenna element. The corresponding transmit and receive index sets are denoted by
\begin{align}
\mathcal I_{\rm T}&\ =\ \{i_{{\rm T},1},i_{{\rm T},2},\ldots,i_{{\rm T},M_{\rm T}}\},
\\
\mathcal I_{\rm R}&\ =\ \{i_{{\rm R},1},i_{{\rm R},2},\ldots,i_{{\rm R},M_{\rm R}}\}.
\end{align}

Under the far-field approximation, the transmit and receive path
lengths are respectively expressed as
\begin{align}
r_{{\rm T},m}&\approx r_{{\rm T},c}
-i_{{\rm T},m}d_{\rm T}\sin\theta,\\
r_{{\rm R},n}&\approx r_{{\rm R},c}
-i_{{\rm R},n}d_{\rm R}\sin\theta.
\label{eq:farfield_path_lengths}
\end{align}
Then, the corresponding transmit--target--receive delay is given by
\begin{equation}
\tau_{n,m}
=
\frac{r_{{\rm T},m}+r_{{\rm R},n}}{c_0}
\approx
\tau-\frac{d_{n,m}\sin\theta}{c_0},
\label{eq:channel_delay}
\end{equation}
where $\tau\triangleq 2r/c_0$ is the reference-channel
round-trip delay with $c_0$ denoting the speed of light, and
$d_{n,m}\triangleq i_{{\rm T},m}d_{\rm T}
+i_{{\rm R},n}d_{\rm R}$ is the effective displacement of the
$(m,n)$ channel.

Based on \eqref{eq:channel_delay}, the passband signal received by the $n$th element is denoted by
\begin{equation}
y_n(t) = \sum_{m=1}^{M_{\rm T}} \alpha\, \psi_m\bigl(t-\tau_{n,m}\bigr)
e^{j2\pi (f_c+\Delta f_m)(t-\tau_{n,m})}
+ w_n(t),
\end{equation}
where $w_n(t)$ is a zero-mean proper complex white Gaussian noise
process with two-sided power spectral density $N_0$. The noise processes
at different receive elements are assumed independent. The point-target
scattering coefficient $\alpha$ is modeled as an unknown complex scalar
that remains constant within the coherent processing interval
\cite{KayStat1998,Wang2014CRLB,Xu2015TSP}.
A conditional-average FIM interpretation for a sample-varying
scattering coefficient is given in Appendix~\ref{sec:fast_extension}.

After downconversion by $e^{-j2\pi f_ct}$, the received signal
passes through an ideal unity-gain low-pass filter. Its complex-baseband
passband is $\mathcal{B}_{\rm rx}=[-B_{\rm rx}/2,B_{\rm rx}/2]$, where
$B_{\rm rx}$ is the total two-sided bandwidth. We require
\begin{equation}
\mathcal{B}_{\rm agg}\subseteq\mathcal{B}_{\rm rx},
\qquad
B_{\rm rx}\ge B_{\rm agg}.
\end{equation}
Under this ideal receiver assumption, the filtered baseband signal
can be approximated as
\begin{align}
\tilde{y}_n(t)
&\approx
\sum_{m=1}^{M_{\rm T}}
\alpha\,s_{n,m}(t)
+\tilde{w}_n(t),
\label{eq:baseband_received}
\end{align}
where $s_{n,m}(t)$ denotes the deterministic response of the
$(m,n)$ channel excluding the scattering coefficient, given by
\begin{align}
s_{n,m}(t)
&=\psi_m(t-\tau_{n,m})e^{j2\pi\Delta f_m(t-\tau)}\nonumber\\
&\quad\times e^{j2\pi f_c d_{n,m}\sin\theta/c_0},
\label{eq:snm}
\end{align}
and $\tilde{w}_n(t)$ is the filtered noise signal. To obtain
\eqref{eq:snm}, we use the far-field delay approximation and factor the
propagation phase as
\begin{equation*}
\begin{aligned}
e^{-j2\pi(f_c+\Delta f_m)\tau_{n,m}}
&\approx e^{-j2\pi f_c\tau}e^{-j2\pi\Delta f_m\tau}\\
&\quad\times e^{j2\pi(f_c+\Delta f_m)d_{n,m}\sin\theta/c_0}.
\end{aligned}
\end{equation*}
As $\alpha$ is an unknown complex nuisance parameter, the common
phase $e^{-j2\pi f_c\tau}$ is not separately identifiable from the
phase of $\alpha$ and is therefore absorbed into it.
The remaining FDA--aperture phase
$e^{j2\pi\Delta f_m d_{n,m}\sin\theta/c_0}$ is neglected when
\begin{equation}
|2\pi\Delta f_m d_{n,m}\sin\theta/c_0|\ll 1.
\label{eq:residual FDA--aperture phase}
\end{equation}

The ideal receiver-bandlimited signal is sampled at the complex-baseband
Nyquist rate\footnote{A complete Nyquist-rate sample
sequence represents the ideal bandlimited signal without loss. Here we
use a finite observation interval of duration $T$ and neglect
out-of-window energy and boundary effects.}, $T_s=1/B_{\rm rx}$. The samples are taken at
$t_k=kT_s$, $k=0,\ldots,K-1$, with $K=T/T_s=TB_{\rm rx}$;
for simplicity, $TB_{\rm rx}$ is assumed to be an integer.

The unit-scattering sample at receiver $n$ is given by
\begin{equation}
s_{n,k}=\sum\nolimits_{m=1}^{M_{\rm T}}s_{n,m}(t_k).
\end{equation}
Define
$\mathbf{s}_n
=[s_{n,0},s_{n,1},\ldots,s_{n,K-1}]^{\rm T}$,
and collect the receive channels as
$\mathbf{s}
=[\mathbf{s}_1^{\rm T},\ldots,\mathbf{s}_{M_{\rm R}}^{\rm T}]^{\rm T}$.
The discrete observation model is therefore
\begin{equation}
\mathbf{y} = \alpha\, \mathbf{s} + \mathbf{w}, \label{eq:discrete_received_signal}
\end{equation}
where $\mathbf w$ is the noise vector.
For the ideal rectangular receiver response, the filtered noise
autocorrelation at each receive element is
\begin{equation}
R_{\widetilde w}(\tau)
=
N_0B_{\rm rx}
{\rm sinc}(B_{\rm rx}\tau),
\end{equation}
where ${\rm sinc}(x)=\sin(\pi x)/(\pi x)$. Consequently, with
$w_n[k]\triangleq\tilde w_n(t_k)$,
\begin{equation}
\mathbb E\{w_n[k]w_{n'}^*[\ell]\}
=
N_0B_{\rm rx}\delta_{nn'}\delta_{k\ell}.
\end{equation}
Hence, the stacked noise vector satisfies
$\mathbf w\sim\mathcal{CN}(\mathbf 0,\sigma^2\mathbf I)$ with
$\sigma^2=N_0B_{\rm rx}$.

\begin{remark}
The observation model in \eqref{eq:discrete_received_signal} is formed
without an initial reduction to transmit-waveform matched-filter outputs.
It therefore retains the explicit dependence on both the delayed
waveform envelopes and the FDA-induced phases for the subsequent FIM
calculation.
\end{remark}

\section{Unified CRLB Formulation}
\label{sec:modelII}
Based on the discrete observation model in
\eqref{eq:discrete_received_signal}, we derive the CRLB for joint
range--angle estimation in the presence of an unknown deterministic
complex scattering coefficient. We first construct the FIM for all
unknown parameters and then eliminate the scattering coefficient by a
Schur complement. For each fixed receiver configuration, the noise
variance $\sigma^2=N_0B_{\rm rx}$ is assumed known and independent of
the unknown parameters.

Let $\alpha=\alpha_{\rm R}+j\alpha_{\rm I}$ and define the real-valued
parameter vector
\begin{equation}
\boldsymbol{\eta}
\triangleq
\begin{bmatrix}
r & \theta & \alpha_{\rm R} & \alpha_{\rm I}
\end{bmatrix}^{\rm T}.
\label{eq:parameter_vector}
\end{equation}
The observation vector $\mathbf y\in\mathbb C^{KM_{\rm R}}$ has
mean $\boldsymbol\mu=\alpha\mathbf s(r,\theta)$ and covariance
$\sigma^2\mathbf I$. Its likelihood is
\begin{equation}
p(\mathbf y;\boldsymbol\eta)
=
\frac{1}{(\pi\sigma^2)^{KM_{\rm R}}}
\exp\left(
-\frac{\|\mathbf y-\alpha\mathbf s\|_2^2}{\sigma^2}
\right).
\label{eq:likelihood_modelII}
\end{equation}
Since the covariance is independent of $\boldsymbol\eta$, the
$(i,j)$th entry of the FIM is \cite{KayStat1998}
\begin{equation}
J_{ij}
=
\frac{2}{\sigma^2}
\Re\!\left\{
\left(\frac{\partial\boldsymbol\mu}{\partial\eta_i}\right)^{\rm H}
\left(\frac{\partial\boldsymbol\mu}{\partial\eta_j}\right)
\right\}.
\label{eq:fim_formula_modelII}
\end{equation}
Defining $\mathbf s_r\triangleq\partial\mathbf s/\partial r$ and
$\mathbf s_\theta\triangleq\partial\mathbf s/\partial\theta$, the
required derivatives are
$\partial\boldsymbol\mu/\partial r=\alpha\mathbf s_r$,
$\partial\boldsymbol\mu/\partial\theta=\alpha\mathbf s_\theta$,
$\partial\boldsymbol\mu/\partial\alpha_{\rm R}=\mathbf s$, and
$\partial\boldsymbol\mu/\partial\alpha_{\rm I}=j\mathbf s$.

Let $\mathbf p=[r,\theta]^{\rm T}$ denote the parameters of interest
and $\boldsymbol a=[\alpha_{\rm R},\alpha_{\rm I}]^{\rm T}$ the nuisance
parameters. Partition the FIM according to
$\boldsymbol\eta=[\mathbf p^{\rm T},\boldsymbol a^{\rm T}]^{\rm T}$ as
\begin{equation}
\mathbf J(\boldsymbol\eta)
=
\begin{bmatrix}
\mathbf J_{pp} & \mathbf J_{pa}\\
\mathbf J_{ap} & \mathbf J_{aa}
\end{bmatrix}.
\label{eq:fim_partition}
\end{equation}
To make the nuisance-parameter elimination explicit, define
$\mathbf D=[\mathbf s_r,\mathbf s_\theta]$. The relevant diagonal
blocks are
\[
\mathbf J_{pp}
=\frac{2|\alpha|^2}{\sigma^2}
\Re\{\mathbf D^{\rm H}\mathbf D\},\qquad
\mathbf J_{aa}
=\frac{2\mathbf s^{\rm H}\mathbf s}{\sigma^2}\mathbf I_2.
\]
For $\mathbf s\ne\mathbf0$, $\mathbf J_{aa}$ is invertible. Treating
$\boldsymbol a$ as a nuisance parameter, the equivalent FIM (EFIM)
for $\mathbf p$ follows from the Schur complement \cite{Xu2015TSP}:
\begin{equation}
\mathbf J_{\rm e}(\mathbf p)
=
\mathbf J_{pp}-\mathbf J_{pa}\mathbf J_{aa}^{-1}\mathbf J_{ap}
=|\alpha|^2\mathbf G(\mathbf p),
\label{eq:FIM_projection}
\end{equation}
where $\mathbf G(\mathbf p)$ is independent of the unknown complex
scattering coefficient and is given by
\begin{equation}
\mathbf G(\mathbf p)
=
\frac{2}{\sigma^2}
\begin{bmatrix}
\Re\{\mathbf s_r^{\rm H}\mathbf P_{\mathbf s}^{\perp}\mathbf s_r\}
&
\Re\{\mathbf s_r^{\rm H}\mathbf P_{\mathbf s}^{\perp}\mathbf s_\theta\}
\\
\Re\{\mathbf s_\theta^{\rm H}\mathbf P_{\mathbf s}^{\perp}\mathbf s_r\}
&
\Re\{\mathbf s_\theta^{\rm H}\mathbf P_{\mathbf s}^{\perp}\mathbf s_\theta\}
\end{bmatrix}.
\label{eq:G_def_modelII}
\end{equation}
Here,
$\mathbf P_{\mathbf s}^{\perp}
=\mathbf I-\mathbf s\mathbf s^{\rm H}/(\mathbf s^{\rm H}\mathbf s)$
is the orthogonal projector onto the complement of
$\operatorname{span}\{\mathbf s\}$. It removes derivative components
parallel to $\mathbf s$, which cannot be distinguished from changes
in the unknown complex scattering coefficient. 

If $\alpha\ne0$ and $\mathbf G(\mathbf p)$ is nonsingular, the
CRLB matrix for $\mathbf p$ is
\begin{equation}
\mathrm{CRLB}(\mathbf p)
=\mathbf J_{\rm e}^{-1}(\mathbf p)
=\frac{1}{|\alpha|^2}\mathbf G^{-1}(\mathbf p).
\label{eq:crb_unified}
\end{equation}
Equation~\eqref{eq:crb_unified} provides the general CRLB for the
adopted sampled-signal model. The waveform-delay and FDA phase
dependencies remain explicit through $\mathbf s_r$ and
$\mathbf s_\theta$. Section~\ref{sec:CF CRLB} develops tractable
closed-form approximations under additional waveform and array
conditions.




\section{Closed-Form CRLB Representation}
\label{sec:CF CRLB}
The general EFIM in \eqref{eq:G_def_modelII} involves
nuisance-parameter projection corrections in both its diagonal and
off-diagonal entries. It also does not explicitly reveal the dependence
on waveform RMS bandwidth, FDA frequency offsets, and array geometry.
This section derives a tractable closed-form approximation.
We first specify the waveform and array conditions, then simplify
the projection terms under these conditions, and finally evaluate
the range, angle, and cross-information terms.

\subsection{Signal and Array Conditions}
\label{sec:common}
This subsection presents the waveform and array assumptions needed to
obtain tractable closed-form CRLBs from the general EFIM in
\eqref{eq:G_def_modelII}.

Let $\Psi_m(f)$ be the Fourier transform of $\psi_m(t)$, and let
$E_m=\int|\psi_m(t)|^2dt=\int|\Psi_m(f)|^2df$ for the $m$th baseband waveform. The spectral centroid $\bar f_m$ and
the RMS bandwidth $\beta_m$ are defined respectively as \cite{oppenheim_signals,KayStat1998}
\begin{align}
\bar f_m &\triangleq \frac{1}{E_m}\int f|\Psi_m(f)|^2df,\\
\beta_m^2 &\triangleq \frac{1}{E_m}\int (f-\bar f_m)^2|\Psi_m(f)|^2df.
\label{eq:beta}
\end{align}

We assume that each baseband waveform is spectrally centered,
i.e.,
$\bar f_m= 0,\ m=1,\ldots,M_{\rm T}.$
Then, by Parseval's theorem, one has
\begin{align}
\int |\psi_m'(t)+j2\pi f_0\psi_m(t)|^2dt
&=
(2\pi)^2E_m\left(\beta_m^2+f_0^2\right),
\label{eq:q_identity}
\end{align}
where $f_0\in\mathbb R$ is an arbitrary frequency shift, and
\begin{equation}
    \begin{cases}
\int \psi_m'^{*}(t)\psi_m(t)\, dt
=
-j2\pi E_m\bar{f}_m = 0,
\\
\int \psi_m^{*}(t)\psi_m'(t)\, dt =j2\pi E_m\bar{f}_m = 0.
    \end{cases}
\label{eq:cross_zero}
\end{equation}
We further assume equal waveform energy and RMS bandwidth,
i.e., $E_m=E$ and $\beta_m=\beta$ for all transmit elements.

Motivated by the standard waveform-orthogonality assumption in
colocated MIMO and FDA-MIMO radar \cite{Li2007,Sammartino2013TAES,
Xu2015TSP,Xiong2018TAES}, we impose the following local
delay--frequency cross-orthogonality condition to decouple the contributions of different transmit waveforms in the
closed-form derivation:
\begin{equation}
\int
\psi_m^{(a)}(t)
\left[
\psi_\ell^{(b)}
\!\left(t-\Delta\tau_{m\ell}\right)
\right]^*
e^{j2\pi(\Delta f_m-\Delta f_\ell)t}
\,dt
\approx0,
\label{eq:orthogonality}
\end{equation}
where $m\neq\ell$, $a,b\in\{0,1\}$, 
\[
\psi_m^{(0)}(t)\triangleq\psi_m(t),\qquad
\psi_m^{(1)}(t)\triangleq\psi'_m(t),
\]
and \[
\Delta\tau_{m\ell}
=
\tau_{n,\ell}-\tau_{n,m}
=
-\frac{(i_{{\rm T},\ell}-i_{{\rm T},m})d_{\rm T}\sin\theta}{c_0}.
\]
The centered linear
frequency-offset law will be adopted hereafter
\begin{equation}
\Delta f_m=i_{{\rm T},m}\Delta f,
\qquad m=1,\ldots,M_{\rm T},
\label{eq:linear_increment_law}
\end{equation}
as commonly used in FDA-MIMO CRLB studies
\cite{Xu2015TSP,Xiong2018TAES}.

For the centered index sets defined in \eqref{eq:centered_transmit_index} and \eqref{eq:centered_receive_index},
all odd-order moments vanish, i.e.,
\begin{align}
\frac{1}{M_{\rm T}}\sum_{m=1}^{M_{\rm T}}i_{{\rm T},m}^{a}
=
\frac{1}{M_{\rm R}}\sum_{n=1}^{M_{\rm R}}i_{{\rm R},n}^{a}
=0,
\label{eq:center array}
\end{align}
for any positive odd integer $a$. We define their centered second-order moments as
\begin{equation}
    \begin{cases}
    \sigma_{\rm T}^2 \triangleq \frac{1}{M_{\rm T}}\sum_{m=1}^{M_{\rm T}} i_{{\rm T},m}^2,\\
    \sigma_{\rm R}^2 \triangleq \frac{1}{M_{\rm R}}\sum_{n=1}^{M_{\rm R}} i_{{\rm R},n}^2. 
    \end{cases}\label{eq:centered second moment}
\end{equation}
The effective aperture variance is given by
\begin{equation}
\sigma_d^2
\triangleq
\frac{1}{M}\sum_{n=1}^{M_{\rm R}}\sum_{m=1}^{M_{\rm T}} d_{n,m}^2
=
\sigma_{\rm T}^2d_{\rm T}^2+\sigma_{\rm R}^2d_{\rm R}^2 ,
\label{eq:sigma_d_new}
\end{equation}
where $M=M_{\rm T}M_{\rm R}$ is the number of
transmit--receive channels, and
$d_{n,m}=i_{{\rm T},m}d_{\rm T}+i_{{\rm R},n}d_{\rm R}$.

\subsection{Projection-Correction Approximation}
We now evaluate the range projection correction $\lvert\mathbf s^{\rm H}\mathbf s_r\rvert^2$ and
angle projection correction $\lvert\mathbf s^{\rm H}\mathbf s_\theta\rvert^2$ in \eqref{eq:G_def_modelII} under the conditions above. 

\subsubsection{Range Projection Correction}  

Collect the factors independent of $\tau$ as
\begin{equation}
    f_{n,m}^{(r)}(t,\theta)
    \triangleq
    e^{j2\pi \Delta f_m t}e^{j2\pi f_c d_{n,m}\sin\theta/c_0},
\end{equation}
such that
\begin{equation}
    s_{n,m}(t)=
    \psi_m\big(t-\tau_{n,m}\big)e^{-j2\pi \Delta f_m\tau}f_{n,m}^{(r)}(t,\theta).
\end{equation}
Using $\partial\tau/\partial r=2/c_0$ gives
\begin{align}
\frac{\partial s_{n,m}(t)}{\partial r}
=\ &
-\frac{2}{c_0}
 f_{n,m}^{(r)}(t,\theta)e^{-j2\pi \Delta f_m\tau}\notag\\
&\times\big(\psi_m'(t-\tau_{n,m})+j2\pi \Delta f_m\psi_m(t-\tau_{n,m})\big).
\label{eq:ds_dr}
\end{align}

Then, $|\mathbf s^H\mathbf s_r|$ expands as
\begin{align}
    |\mathbf s^H\mathbf s_r| &= \bigg|\sum_{n=1}^{M_{\rm R}}\sum_{k=0}^{K-1}\sum_{m_1=1}^{M_{\rm T}}s_{n,m_1}^*(t_k)\sum_{m_2=1}^{M_{\rm T}}\frac{\partial s_{n,m_2}(t_k)}{\partial r}\bigg|, 
    \label{eq:s_r-s_r_H}\\
    &\overset{(i)}{\approx}\bigg|\sum_{n=1}^{M_{\rm R}}B_{\rm rx}\int \sum_{m_1=1}^{M_{\rm T}}s_{n,m_1}^*(t)\sum_{m_2=1}^{M_{\rm T}}\frac{\partial s_{n,m_2}(t)}{\partial r}\,dt\bigg|
\end{align}
where $t_k$ denotes a sampling time instant, and approximation (i) replaces the finite sample sum
by its continuous-time counterpart,
$\sum_k g(t_k)\approx B_{\rm rx}\int g(t)\,dt$,
assuming negligible observation-window boundary
effects.
Under \eqref{eq:orthogonality}, the cross-waveform terms $\int s_{n,m_1}^*(t)\frac{\partial s_{n,m_2}(t)}{\partial r}\,dt$ for
$m_1\neq m_2$ are negligible, yielding
\begin{align}
    |\mathbf s^H\mathbf s_r| &\ {\approx}\ \bigg|B_{\rm rx}\sum_{n=1}^{M_{\rm R}}\sum_{m=1}^{M_{\rm T}}\int s_{n,m}^*(t)\frac{\partial s_{n,m}(t)}{\partial r}\,dt\bigg| 
    \label{eq:discrete s_r-s_r_H}\\
    &\overset{(i)}{\approx} \ \frac{4\pi}{c_0}  B_{\rm rx} E\bigg|\sum_{n=1}^{M_{\rm R}}\sum_{m=1}^{M_{\rm T}} i_{{\rm T},m}\Delta f\bigg|,
\end{align}
where $(i)$ follows by substituting \eqref{eq:snm} and \eqref{eq:ds_dr} into \eqref{eq:discrete s_r-s_r_H} and applying the zero spectral centroid properties of waveforms in \eqref{eq:cross_zero}.

Finally, using the centered linear offset law
\eqref{eq:linear_increment_law} and the odd-moment condition
\eqref{eq:center array} yields
\begin{align}
    |\mathbf s^H\mathbf s_r|
    &{\approx}0. \label{eq:zero projection s_r}
\end{align}

\subsubsection{Angle Projection Correction}
For the angular derivative, collect the factors independent of
$\theta$ as
\begin{equation}
    f_{n,m}^{(\theta)}(t,\tau)
    \triangleq
    e^{j2\pi \Delta f_m t}e^{-j2\pi \Delta f_m\tau},
\end{equation}
so that
\begin{equation}
    s_{n,m}(t)=
    \psi_m\big(t-\tau_{n,m}\big)e^{j2\pi f_c d_{n,m}\sin\theta/c_0}f_{n,m}^{(\theta)}(t,\tau).
\end{equation}
Using $\partial\tau_{n,m}/\partial\theta=-d_{n,m}\cos\theta/c_0$, we obtain
\begin{align}
\frac{\partial s_{n,m}(t)}{\partial\theta}
=\ &
\frac{d_{n,m}\cos\theta}{c_0}
 f_{n,m}^{(\theta)}(t,\tau)e^{j2\pi f_c d_{n,m}\sin\theta/c_0}\notag\\
&\times\big(\psi_m'(t-\tau_{n,m})+j2\pi f_c\psi_m(t-\tau_{n,m})\big).
\label{eq:ds_dtheta}
\end{align}
Then, analogous to simplifying $|\mathbf s^H\mathbf s_r|$ in \eqref{eq:s_r-s_r_H}-\eqref{eq:zero projection s_r}, we have 
\begin{align}
    |\mathbf s^H\mathbf s_\theta| 
    &\overset{(i)}{\approx} 
    \bigg|B_{\rm rx}\sum_{n=1}^{M_{\rm R}}\sum_{m=1}^{M_{\rm T}}\int s_{n,m}^*(t)\frac{\partial s_{n,m}(t)}{\partial \theta}\,dt\bigg|\notag\\
    &\overset{(ii)}{\approx}
    \bigg|\frac{2\pi}{c_0}  B_{\rm rx} E f_c \cos\theta
    \sum_{n=1}^{M_{\rm R}}\sum_{m=1}^{M_{\rm T}} d_{n,m}\bigg| \notag\\
    &\overset{(iii)}{=}0,
\end{align}
where (i) comes from the waveform orthogonality \eqref{eq:orthogonality}, (ii) is based on  
the zero spectral centroid properties \eqref{eq:cross_zero}, and (iii) is due to $\sum_{n,m}d_{n,m}=0$ under the
centered-array odd-order moment condition \eqref{eq:center array}.

Under the adopted signal and array conditions, the signal--derivative
inner products vanish to leading order, so the diagonal and
cross-projection corrections in \eqref{eq:G_def_modelII} can be
neglected. The scattering-coefficient-independent EFIM factor is
therefore approximated as
$\mathbf G({\bm p})\approx\mathbf G_0({\bm p})$, where
\begin{align}
\mathbf G_0({\bm p})
&=\frac{2}{\sigma^2}
\begin{bmatrix}
\|\mathbf{s}_r\|_2^2
&
\Re\{\mathbf s_r^H\mathbf s_\theta\}
\\
\Re\{\mathbf s_\theta^H\mathbf s_r\}
&
\|\mathbf{s}_\theta\|_2^2
\end{bmatrix}
\\
&\triangleq\begin{bmatrix}
G_{rr} & G_{r\theta}\\
G_{r\theta} & G_{\theta\theta}
\end{bmatrix}.
\label{eq:G_0}
\end{align}
Then, the CRLB in \eqref{eq:crb_unified} can be approximated as
\begin{align}
\mathrm{CRLB}(\mathbf p)
\approx
\frac{1}{|\alpha|^2}\,\mathbf G_0^{-1}({\bm p}).
\label{eq:crb_simplified}
\end{align}
It remains to evaluate the three distinct entries of
$\mathbf G_0(r,\theta)$.

\subsection{Closed-Form CRLBs}
The closed-form CRLB follows by evaluating the range derivative energy
$\|\mathbf s_r\|_2^2$, the cross-derivative inner product
$\Re\{\mathbf s_r^{\rm H}\mathbf s_\theta\}$, and
the angle derivative energy $\|\mathbf s_\theta\|_2^2$.

\subsubsection{Range Derivative Energy}
Combining \eqref{eq:q_identity} and \eqref{eq:ds_dr}, we have the range derivative integral
\begin{equation}
\int \left|\frac{\partial s_{n,m}(t)}{\partial r}\right|^2dt
=
\left(\frac{2}{c_0}\right)^2(2\pi)^2E\big(\beta^2+i_{{\rm T},m}^2\Delta f^2\big).
\label{eq:dr_energy}
\end{equation}
The equal-energy and equal-bandwidth conditions have been used above.
\eqref{eq:dr_energy} shows that the range derivative energy per channel contains
two distinct contributions, i.e., the waveform-delay contribution
governed by $\beta^2$ and the FDA phase contribution governed
by the element-dependent frequency offset.

Analogous to evaluating $|\mathbf s^H\mathbf s_r|$, applying \eqref{eq:orthogonality} and the sampling-to-integral
approximation yields
\begin{align}
    \|\mathbf{s}_r\|_2^2
    &\ {\approx}
    \sum_{n=1}^{M_{\rm R}}\sum_{m=1}^{M_{\rm T}}B_{\rm rx}
    \int\left|\frac{\partial s_{n,m}(t)}{\partial r}\right|^2dt\notag\\
    &=
M_{\rm R}\left(\frac{4\pi}{c_0}\right)^2 B_{\rm rx}
\sum_{m=1}^{M_{\rm T}} E(\beta^2+i_{{\rm T},m}^2\Delta f^2)
\notag\\
&=
\left(\frac{4\pi}{c_0}\right)^2M B_{\rm rx} E
\left(\beta^2+B_{\rm F}^2\right),
\label{eq:sr_norm_new}
\end{align}
where 
\begin{equation}
B_{\rm F}^2 \triangleq \sigma_{\rm T}^2\Delta f^2.
\label{eq:BF_new}
\end{equation}
Under \eqref{eq:linear_increment_law},
$B_{\rm F}$ equals the RMS spread of the transmit-element
frequency offsets and is therefore referred to as the equivalent
FDA frequency-diversity bandwidth.

\subsubsection{Angle Derivative Energy}
Likewise, applying \eqref{eq:q_identity} to \eqref{eq:ds_dtheta} yields the angle derivative energy per channel as
\begin{equation}
\int \left|\frac{\partial s_{n,m}(t)}{\partial\theta}\right|^2dt
=
\left(\frac{d_{n,m}\cos\theta}{c_0}\right)^2
(2\pi)^2E\big(\beta^2+f_c^2\big),
\label{eq:dtheta_energy}
\end{equation}
which depends on
$d_{n,m}$, $\beta$, and $f_c$.

Similarly, one has
\begin{align}
\|\mathbf s_\theta\|_2^2
&\ {\approx}\
\sum_{n=1}^{M_{\rm R}}\sum_{m=1}^{M_{\rm T}}B_{\rm rx}
\int \left|\frac{\partial s_{n,m}(t)}{\partial \theta}\right|^2\, dt\notag\\
&\ {\approx}\
\left(\frac{2\pi}{c_0}\right)^2 B_{\rm rx}
\sum_{n,m} d_{n,m}^2 \cos^2\theta\,E(\beta^2+f_c^2)
\notag\\
& \ {=}\
\left(\frac{2\pi}{c_0}\right)^2M B_{\rm rx} E
(\beta^2+f_c^2)\sigma_d^2\cos^2\theta,
\label{eq:st_norm_new}
\end{align}
where $\sum_{n,m} d_{n,m}^2=M\sigma_d^2$ has been used based on \eqref{eq:sigma_d_new}.

\subsubsection{Cross-Derivative Inner Product}
Applying the waveform orthogonality \eqref{eq:orthogonality} and the sampling-to-integral
approximation, one has the cross-derivative inner product
\begin{align}
\mathbf s_r^H\mathbf s_\theta
    &\ {\approx}\
    \sum_{n=1}^{M_{\rm R}}\sum_{m=1}^{M_{\rm T}}B_{\rm rx}
    \int \left(\frac{\partial s_{n,m}(t)}{\partial r}\right)^{\!*}
    \left(\frac{\partial s_{n,m}(t)}{\partial \theta}\right)\,dt\notag\\
    &\overset{(i)}{\approx} \
    -2B_{\rm rx}\cos\theta\left(\frac{2\pi}{c_0}\right)^2
    \sum_{n,m}d_{n,m} E \left(\beta^2 + i_{{\rm T},m}\Delta f\,f_c\right)\notag\\
    &\overset{(ii)}{=}\
    -\frac{8\pi^2}{c_0^2} M B_{\rm rx} E
    \,(f_c\Delta f)\,\sigma_{\rm T}^2\,d_{\rm T}\,\cos\theta,
\label{eq:srst_new}
\end{align}
where approximation (i) is obtained by substituting the derivative expressions in
\eqref{eq:ds_dr} and \eqref{eq:ds_dtheta}, and (ii) is because the term proportional to $\beta^2$ vanishes as $\sum_{n,m}d_{n,m}=0$ under the centered-array condition \eqref{eq:center array}.

\subsubsection{Closed-Form Bounds}
According to \eqref{eq:G_0}-\eqref{eq:crb_simplified}, the range and angle CRLB are given by
\begin{align}
\mathrm{CRLB}_r &= \frac{1}{|\alpha|^2}\frac{1}{G_{rr}(1-\rho^2)},\label{eq:CRLB_r general}\\
\mathrm{CRLB}_\theta &= \frac{1}{|\alpha|^2}\frac{1}{G_{\theta\theta}(1-\rho^2)},\label{eq:CRLB_t general}
\end{align}
where $\rho^2$ is the normalized cross-information factor, defined as
\begin{equation}
\rho^2 \triangleq \frac{(G_{r\theta})^2}{G_{rr}G_{\theta\theta}}.
\label{eq:rho_def_new}
\end{equation}
For a nonsingular FIM, $0\le\rho^2<1$ is also called the range--angle coupling factor in this paper.

Substituting \eqref{eq:sr_norm_new}, \eqref{eq:st_norm_new} and \eqref{eq:srst_new} into \eqref{eq:G_0}, and using $\sigma^2=N_0B_{\rm rx}$, we have
\begin{align}
G_{rr}
&=
2\gamma_E
M\left(\frac{4\pi}{c_0}\right)^2
\left(\beta^2+B_{\rm F}^2\right),
\label{eq:Jrr0}
\\
G_{\theta\theta}
&=
2\gamma_E
M\left(\frac{2\pi}{c_0}\right)^2
(\beta^2+f_c^2)\sigma_d^2\cos^2\theta,
\label{eq:Jtt0}
\\
G_{r\theta}
&=
-16\gamma_EM\frac{\pi^2}{c_0^2}
(f_c\Delta f)\,d_{\rm T}\,\sigma_{\rm T}^2\cos\theta,
\label{eq:Jrt0}
\end{align}
where $\gamma_E=E/N_0$ is the energy-to-noise-density ratio (ENR). 
Then, the coupling factor $\rho^2$ can be factorized as the product of three coupling coefficients, i.e.
\begin{equation}
{
\rho^2 = \mu_a\,\mu_c\,\mu_{\rm F}.
}
\label{eq:rho_product_new}
\end{equation}
The geometric coupling coefficient is defined as
\begin{equation}
\mu_a
\triangleq
\frac{\sigma_{\rm T}^2 d_{\rm T}^2}{\sigma_d^2},
\qquad
0\le\mu_a\le1,
\label{eq:mua_new}
\end{equation}
which measures the fraction of the effective aperture variance contributed by the transmit array.
The carrier--bandwidth coupling coefficient is given by
\begin{equation}
\mu_c
\triangleq
\frac{f_c^2}{f_c^2+\beta^2},
\qquad
0<\mu_c\le 1,
\label{eq:muc_new}
\end{equation}
which measures the relative dominance of the carrier frequency term over waveform RMS bandwidth.
The FDA--bandwidth coupling coefficient is
\begin{equation}
\mu_{\rm F}
\triangleq
\frac{B_{\rm F}^2}{B_{\rm F}^2+\beta^2},
\qquad
0\le \mu_{\rm F}\le 1,
\label{eq:muF_new}
\end{equation}
which quantifies the relative contribution of FDA frequency-diversity
bandwidth relative to waveform RMS bandwidth.

Finally, substituting \eqref{eq:Jrr0}--\eqref{eq:Jrt0} into
\eqref{eq:CRLB_r general}--\eqref{eq:CRLB_t general} yields
\begin{align}
{\mathrm{CRLB}_r}
&=
\frac{1}{|\alpha|^2}\,
\frac{c_0^2}{32\pi^2 M \gamma_E}
\frac{1}{I_r},
\label{eq:CRLBr_unified}
\\
{\mathrm{CRLB}_\theta}
&=
\frac{1}{|\alpha|^2}\,
\frac{c_0^2}{8\pi^2 M \gamma_E}
\frac{1}{I_\theta},
\label{eq:CRLBtheta_unified}
\end{align}
where
$I_r$ is the range information
\begin{equation}
I_r
\triangleq
\beta^2+B_{\rm F}^2(1-\mu_a\mu_c),
\label{eq:Ir}
\end{equation}
and $I_\theta$ is the angle information 
\begin{equation}
I_\theta
\triangleq
\cos^2\theta\sigma_d^2\Big[\beta^2+f_c^2(1-\mu_a\mu_{\rm F})\Big].
\label{eq:Itheta}
\end{equation}
Equations~\eqref{eq:Ir} and \eqref{eq:Itheta} capture the dependence of the CRLBs on RMS bandwidth, FDA frequency diversity, carrier frequency, and array aperture. Their implications are further analyzed in Section~\ref{sec:CRLB_analysis}.

\begin{remark}[Special cases]
When the FDA frequency offset vanishes, i.e., $\Delta f=0$, with $\beta>0$, one has
$B_{\rm F}=0$, $\mu_{\rm F}=0$, and hence $\rho^2=0$. The proposed closed-form CRLBs reduce to
\begin{equation}
\mathrm{CRLB}_r^{\mathrm{MIMO}}
=
\frac{1}{|\alpha|^2}
\frac{c_0^2}{32\pi^2 M\gamma_E}
\frac{1}{\beta^2},
\end{equation}
and
\begin{equation}
\mathrm{CRLB}_\theta^{\mathrm{MIMO}}
=
\frac{1}{|\alpha|^2}
\frac{c_0^2}{8\pi^2 M\gamma_E}
\frac{1}{\cos^2\theta\,\sigma_d^2(\beta^2+f_c^2)}.
\end{equation}
Thus, for $\beta\ll f_c$, the range bound \eqref{eq:CRLBr_unified} retains the conventional inverse-RMS-bandwidth dependence
\cite{Li2008TSPRangeCompression}, whereas the angular bound \eqref{eq:CRLBtheta_unified} is dominated
by the carrier--aperture product $f_c^2\sigma_d^2$, consistent with colocated MIMO
array-performance analyses \cite{boyer2010performance}.
Additionally, in the idealized narrowband (NB) FDA limit
$\beta\rightarrow0$ with $B_{\rm F}>0$, one has
$\mu_c\rightarrow1$, $\mu_{\rm F}\rightarrow1$, and
$\rho^2\rightarrow\mu_a$. The proposed closed-form CRLBs reduce to 
\begin{equation}
\mathrm{CRLB}_r^{\mathrm{NB}}
=
\frac{1}{|\alpha|^2}
\frac{c_0^2}{32\pi^2 M\gamma_E}
\frac{1}{B_{\rm F}^2(1-\mu_a)}, \label{eq:NB FDA-MIMO range}
\end{equation}
and
\begin{equation}
\mathrm{CRLB}_\theta^{\mathrm{NB}}
=
\frac{1}{|\alpha|^2}
\frac{\lambda_c^2}{8\pi^2 M\gamma_E}
\frac{1}{\cos^2\theta\,\sigma_d^2(1-\mu_a)}, \label{eq:NB FDA-MIMO angle}
\end{equation}
where $\lambda_c$ is the carrier wavelength. The resulting dependence on $B_{\rm F}$, the effective aperture,
and the geometric factor $1-\mu_a$ is consistent with existing
NB FDA-MIMO CRLB results
\cite{Xu2015TSP,Xiong2018TAES,Lan2025SPLFreqIncOpt}.
\end{remark}


\section{CRLB Parameter Dependence and Spectral Constraints}
\label{sec:CRLB_analysis}

The closed-form bounds in \eqref{eq:CRLBr_unified}--\eqref{eq:CRLBtheta_unified}
show that the range and angle CRLBs are governed by the information terms
$I_r$ and $I_\theta$, respectively. Based on this, this section analyzes the CRLB dependence on
the waveform RMS bandwidth, FDA frequency diversity, carrier frequency, array
geometry, and target angle, as well as the associated range--angle coupling.
Subsequently, the joint allocation of waveform bandwidth and FDA frequency span is examined under a finite aggregate spectral constraint.

\subsection{Range--Angle Coupling}
From \eqref{eq:rho_product_new}, the squared normalized
cross-information factor is
$\rho^2=\mu_a\mu_c\mu_{\rm F}$
which comprises geometric,
carrier--bandwidth, and FDA--bandwidth components.

For the practically relevant regime $\beta\ll f_c$, one has
$\mu_c\approx1$. The geometric factor $\mu_a$ is determined by
the relative transmit and receive aperture contributions, whereas
$\mu_F$ is controlled by the ratio between the FDA frequency
spread and the waveform RMS bandwidth. Hence, increasing
$\Delta f$ strengthens the range--angle coupling through
$\mu_F$, while increasing $\beta$ weakens it. In the FDA-dominant regime $B_{\rm F}/\beta\to\infty$,
the coupling factor approaches $\mu_a\mu_c$, which
further reduces to approximately $\mu_a$ when
$\beta\ll f_c$ \cite{Xu2015TSP}.

\subsection{Range Information}
\label{subsec:range_information}
The effective range information in \eqref{eq:Ir}
consists of the waveform-delay contribution $\beta^2$
and an FDA-induced contribution
$B_{\rm F}^2(1-\mu_a\mu_c)$.
The latter is reduced by range--angle coupling,
whose magnitude depends on waveform RMS bandwidth
and array geometry.

For a fixed RMS bandwidth $\beta$, $\mu_c$ is independent of $\Delta f$ and
$B_F^2=\sigma_T^2\Delta f^2$. Thus,
\begin{equation}
\frac{\partial I_r}{\partial\Delta f}
=
2\sigma_T^2\Delta f\left(1-\mu_a\mu_c\right).
\label{eq:dIr_ddf}
\end{equation}
For $\Delta f>0$ and $\mu_a\mu_c<1$,
\begin{equation}
\frac{\partial I_r}{\partial\Delta f}>0
\implies
\frac{\partial \mathrm{CRLB}_r}{\partial\Delta f}<0.
\label{eq:range_df_monotonic}
\end{equation}
Therefore, increasing the FDA frequency offset monotonically improves the range
CRLB when the remaining parameters are fixed.

For a fixed $\Delta f$, using
$\mu_c=f_c^2/(f_c^2+\beta^2)$ gives
\begin{equation}
I_r
=
\beta^2+B_F^2
\left(
1-\frac{\mu_a f_c^2}{f_c^2+\beta^2}
\right),
\end{equation}
with derivative
\begin{equation}
\frac{\partial I_r}{\partial\beta}
=
2\beta
\left[
1+
\frac{\mu_a B_F^2 f_c^2}
{\left(f_c^2+\beta^2\right)^2}
\right].
\label{eq:dIr_dbeta}
\end{equation}
Hence, for $\beta>0$,
\begin{equation}
\frac{\partial I_r}{\partial\beta}>0
\implies
\frac{\partial \mathrm{CRLB}_r}{\partial\beta}<0.
\label{eq:range_beta_monotonic}
\end{equation}
The RMS bandwidth therefore improves range estimation through the direct increase
in delay sensitivity and the simultaneous reduction of the coupling contribution.
Consequently, $\beta$ and $\Delta f$ both increase the range information when varied
independently while their competition arises only under a finite spectral constraint, as
discussed in Section~\ref{subsec:spectral_tradeoff}.

\subsection{Angle Information}
\label{subsec:angle_information}

For fixed $\beta>0$, increasing $\Delta f$ increases
$B_F^2=\sigma_T^2\Delta f^2$ and, consequently,
$\mu_{\rm F}=B_F^2/(B_F^2+\beta^2)$. From \eqref{eq:Itheta},
\begin{equation}
\frac{\partial I_\theta}{\partial\Delta f}<0
\implies
\frac{\partial \mathrm{CRLB}_\theta}{\partial\Delta f}>0,
\qquad \Delta f>0.
\label{eq:angle_df_monotonic}
\end{equation}
Thus, increasing FDA frequency diversity reduces the angle information through the
stronger range--angle coupling.

For fixed $\Delta f$,
\begin{equation}
I_\theta
=
\sigma_d^2\cos^2\theta
\left[
\beta^2
+
f_c^2
\left(
1-\frac{\mu_a B_F^2}{B_F^2+\beta^2}
\right)
\right],
\end{equation}
and
\begin{equation}
\frac{\partial I_\theta}{\partial\beta}
=
2\beta\sigma_d^2\cos^2\theta
\left[
1+
\frac{\mu_a f_c^2B_F^2}
{\left(B_F^2+\beta^2\right)^2}
\right].
\label{eq:dItheta_dbeta}
\end{equation}
Therefore, for $\beta>0$,
\begin{equation}
\frac{\partial I_\theta}{\partial\beta}>0
\implies
\frac{\partial \mathrm{CRLB}_\theta}{\partial\beta}<0.
\label{eq:angle_beta_monotonic}
\end{equation}
A larger RMS bandwidth increases the angle information both through the direct
$\beta^2$ term and through the reduction of $\mu_{\rm F}$.

When the element spacings scale with the carrier wavelength as
\begin{equation}
d_T=\kappa_T\lambda_c,
\qquad
d_R=\kappa_R\lambda_c,
\label{eq:wavelength_normalized_spacing}
\end{equation}
the effective aperture variance satisfies $\sigma_d^2\propto f_c^{-2}$.
Accordingly, the product $\sigma_d^2f_c^2$ is independent of $f_c$ for a fixed
electrical array geometry. In addition, when $\beta\ll f_c$, $\mu_c\approx1$ and
the contribution proportional to $\sigma_d^2\beta^2$ is negligible. Under these
conditions, both the range and angle CRLBs exhibit only weak dependence on the
carrier frequency.

The target-angle dependence follows directly from \eqref{eq:Itheta}. For fixed
remaining parameters,
\begin{equation}
\mathrm{CRLB}_\theta(\theta)
=
\mathrm{CRLB}_\theta(0)\sec^2\theta.
\label{eq:theta_scaling}
\end{equation}
Hence, the angle CRLB is minimized at broadside and increases toward endfire,
whereas the range CRLB is independent of $\theta$ under the closed-form model.

\subsection{Array-Size Dependence}
\label{subsec:array_size}

For the centered ULA indices defined in Section~IV, the transmit and receive index
variances are
\begin{equation}
\sigma_T^2
=
\frac{M_T^2-1}{12},
\qquad
\sigma_R^2
=
\frac{M_R^2-1}{12}.
\label{eq:index_variance_array}
\end{equation}
Consequently,
\begin{equation}
B_F^2
=
\frac{M_T^2-1}{12}\Delta f^2,
\label{eq:BF_array}
\end{equation}
and
\begin{equation}
\sigma_d^2
=
\frac{M_T^2-1}{12}d_T^2
+
\frac{M_R^2-1}{12}d_R^2.
\label{eq:sigmad_array}
\end{equation}

The following comparisons hold for fixed per-transmit-waveform
energy $E$, element spacings, $\beta$, and frequency increment
$\Delta f$.
Increasing $M_T$ increases the number of transmit--receive channels
$M=M_TM_R$, the FDA frequency-spread term $B_F^2$, and the effective aperture
$\sigma_d^2$, and therefore generally increases both range and angle information.
For a fixed frequency offset, it also increases the FDA frequency span and
hence the aggregate spectral support required by the transmitter--receiver system.

Increasing $M_R$ increases the number of observation channels and the receive-side
contribution to $\sigma_d^2$, while decreasing the geometric coupling factor
$\mu_a$. Since $M_R$ does not affect the FDA frequency span, receive-array
expansion improves both CRLBs and reduces range--angle coupling without increasing
the required FDA spectral support.

\subsection{Summary of Parameter Effects}
\label{subsec:parameter_summary}

The main consequences of the preceding analysis are summarized as follows.
\begin{itemize}
    \item Increasing either $\Delta f$ or $\beta$ increases the range information
    when these two parameters are varied independently. Hence, both reduce
    $\mathrm{CRLB}_r$.

    \item Increasing $\Delta f$ strengthens the FDA--bandwidth coupling and
    increases $\mathrm{CRLB}_\theta$, whereas increasing $\beta$ weakens the
    coupling and reduces $\mathrm{CRLB}_\theta$.

    \item When the array spacings scale with $\lambda_c$ and $\beta\ll f_c$,
    the carrier-frequency dependence of both CRLBs is weak. The angle CRLB follows
    the $\sec^2\theta$ dependence in \eqref{eq:theta_scaling}.

    \item Increasing either array size generally improves the estimation bounds.
    Only an increase in $M_T$, however, enlarges the FDA frequency span and the
    corresponding aggregate spectral requirement.
\end{itemize}
Table~\ref{tab:parameter_effects} summarizes the corresponding closed-form trends.

\begin{table}[!t]
\centering
\caption{Summary of closed-form parameter trends under
independent parameter variations.
Both carrier-frequency trends assume wavelength-normalized
spacings and $\beta\ll f_c$.}
\label{tab:parameter_effects}
\begin{tabular}{c c c c}
\hline
Parameter & Main dependence & $\mathrm{CRLB}_r$ & $\mathrm{CRLB}_\theta$\\
\hline
$M_T$ &
$M,B_F^2,\sigma_d^2\uparrow$ &
typically $\downarrow$ &
typically $\downarrow$
\\
$M_R$ &
$M,\sigma_d^2\uparrow,\ \mu_a\downarrow$ &
$\downarrow$ &
$\downarrow$
\\
$f_c$ &
electrical aperture, $\mu_c$ &
nearly neutral &
nearly neutral
\\
$\Delta f$ &
$B_F^2,\mu_{\rm F}\uparrow$ &
$\downarrow$ &
$\uparrow$
\\
$\beta$ &
$\beta^2$ $\uparrow$, $\mu_c,\mu_{\rm F}\downarrow$ &
$\downarrow$ &
$\downarrow$
\\
$|\theta|$ &
$\cos^2\theta\downarrow$ &
-- &
$\uparrow$
\\
$E/N_0$ &
information scale $\uparrow$ &
$\downarrow$ &
$\downarrow$
\\
\hline
\end{tabular}
\end{table}

\subsection{Spectral-Resource Tradeoffs}
\label{subsec:spectral_tradeoff}
The preceding results reveal two related tradeoffs. 
Increasing $\Delta f$ improves range estimation through the
FDA-induced range-phase information, but simultaneously increases
the range--angle coupling and reduces the angular
information. Hence, FDA frequency diversity introduces a
range--angle performance tradeoff.

The second tradeoff arises when the waveform bandwidth and FDA
frequency span are constrained by a common spectral budget.
Let $B_{\rm w}(\beta)$ denote the occupied bandwidth of a waveform
with an RMS bandwidth $\beta$. Under the centered linear
frequency-offset law, the aggregate occupied bandwidth is
\begin{equation}
B_{\rm agg}(\beta,\Delta f)
=
B_{\rm w}(\beta)+L\Delta f,
\qquad
L\triangleq M_T-1.
\label{eq:Bagg_general}
\end{equation}
For analytical characterization, we adopt
$B_{\rm w}(\beta)\approx\delta\beta$ for a specified waveform
family where $\delta$ is evaluated from a zero-intercept fit to the occupied
bandwidth of the adopted waveform family. Then, we impose the active spectral constraint
$\delta\beta+L\Delta f=B_{\rm tot}$.

To characterize the allocation between waveform bandwidth and
FDA frequency diversity, define $\eta\in[0,1]$ as the fraction
of the total spectral budget assigned to the FDA frequency span.
The corresponding parameters are
\begin{equation}
\Delta f(\eta)
=
\frac{\eta B_{\rm tot}}{L},
\qquad
\beta(\eta)
=
\frac{(1-\eta)B_{\rm tot}}{\delta}.
\label{eq:spectral_allocation}
\end{equation}
Thus, $\eta=0$ corresponds to the waveform-bandwidth endpoint,
whereas $\eta=1$ corresponds to the FDA-frequency-span endpoint.

Under the narrowband condition
$B_{\rm tot}\ll f_c$,
we have $\beta(\eta)\ll f_c$ over the entire allocation interval
and hence $\mu_c\simeq1$. Substituting
\eqref{eq:spectral_allocation} into \eqref{eq:Ir} therefore gives
\begin{equation}
I_r(\eta)
=
\frac{B_{\rm tot}^2}{\delta^2}(1-\eta)^2
+
\frac{\sigma_T^2B_{\rm tot}^2}{L^2}
(1-\mu_a)\eta^2.
\label{eq:Ir_eta_nb}
\end{equation}
Its derivative is
\begin{equation}
\frac{dI_r}{d\eta}
=
-\frac{2B_{\rm tot}^2}{\delta^2}(1-\eta)
+
\frac{2\sigma_T^2B_{\rm tot}^2}{L^2}
(1-\mu_a)\eta .
\label{eq:dIr_deta_nb}
\end{equation}
For $\mu_a<1$, \eqref{eq:dIr_deta_nb} has the unique solution
\begin{equation}
\eta_r^\star
=
\frac{L^2}
{L^2+\sigma_T^2\delta^2(1-\mu_a)}.
\label{eq:eta_range_stationary}
\end{equation}
Moreover,
\begin{equation}
\frac{d^2I_r}{d\eta^2}
=
2B_{\rm tot}^2
\left[
\frac{1}{\delta^2}
+
\frac{\sigma_T^2(1-\mu_a)}{L^2}
\right]
>0.
\label{eq:d2Ir_deta2_nb}
\end{equation}
Hence, $I_r$ decreases for
$0\le\eta<\eta_r^\star$ and increases for
$\eta_r^\star<\eta\le1$. Since the range CRLB is inversely
proportional to $I_r$, $\eta_r^\star$ corresponds to the unique
maximum of the range CRLB under the fixed spectral budget.
The corresponding waveform bandwidth and FDA frequency offset
are
\begin{equation}
\beta^\star
=
\frac{(1-\eta_r^\star)B_{\rm tot}}{\delta},
\qquad
\Delta f^\star
=
\frac{\eta_r^\star B_{\rm tot}}{L}.
\label{eq:range_stationary_parameters}
\end{equation}
Since the interior stationary point minimizes $I_r(\eta)$, the
range-optimal allocation lies at a boundary. Comparing two
endpoints gives
\begin{align}
I_r(1)>I_r(0)
\Longleftrightarrow 
\delta^2\sigma_{\rm T}^2(1-\mu_a)>L^2.
\end{align}
Therefore, the FDA endpoint is range-optimal only when the FDA-information
gain is sufficiently large relative to the waveform-bandwidth
endpoint. Otherwise, the waveform-bandwidth endpoint is preferred.

For the angle information, define
\begin{equation}
A\triangleq\frac{B_{\rm tot}^2}{\delta^2},
\qquad
C\triangleq\frac{\sigma_T^2B_{\rm tot}^2}{L^2}.
\end{equation}
Then
\begin{equation}
\mu_F(\eta)
=
\frac{C\eta^2}
{C\eta^2+A(1-\eta)^2},
\end{equation}
whose derivative is
\begin{equation}
\frac{d\mu_F}{d\eta}
=
\frac{
2AC\eta(1-\eta)
}{
\left[
C\eta^2+A(1-\eta)^2
\right]^2
}
>0,
\qquad 0<\eta<1.
\label{eq:dmuF_deta}
\end{equation}
Accordingly,
\begin{align}
\frac{dI_\theta}{d\eta}
={}&
-2\sigma_d^2\cos^2\theta\,
A(1-\eta)
\nonumber\\
&\times
\left[
1+
\frac{
\mu_a f_c^2C\eta
}{
\left(
C\eta^2+A(1-\eta)^2
\right)^2
}
\right]
<0,
\qquad 0\le\eta<1.
\label{eq:dItheta_deta}
\end{align}
Therefore, the angle information decreases monotonically with
the FDA spectral-allocation ratio, and no interior stationary
point exists for $I_\theta$. Correspondingly,
$\mathrm{CRLB}_\theta$ increases monotonically with $\eta$.
In addition, under $\mu_c\simeq1$,
$\rho^2\simeq\mu_a\mu_F$ also increases monotonically with
$\eta$.

\section{Numerical Validation}
\label{sec:numerical}
\label{sec:numerical_validation}

This section validates the proposed closed-form CRLBs and examines the
parameter dependencies derived in Sections~IV and~V. Unless otherwise
specified, the baseline parameters are listed in
Table~\ref{tab:baseline}. The reported RCRLB is the square root of the corresponding CRLB.
For logarithmic plots, the range RCRLB is normalized by $1$~m and the
angle RCRLB by $1$~rad before applying $10\log_{10}(\cdot)$.

\begin{table}[!t]
\centering
\caption{Common baseline parameters for the numerical experiments.}
\label{tab:baseline}
\begin{tabular}{c c}
\hline
Parameter & Baseline value\\
\hline
Transmit array size & $M_T=5$\\
Receive array size & $M_R=5$\\
Carrier frequency & $f_c=10~\mathrm{GHz}$\\
Carrier wavelength & $\lambda_c=c_0/f_c$\\
Element spacing & $d_T=d_R=\lambda_c/2$\\
Target range & $r_0=1000~\mathrm{m}$\\
Target angle & $\theta_0=20^\circ$\\
FDA frequency increment & $\Delta f=2~\mathrm{MHz}$\\
Waveform RMS bandwidth & $\beta=4~\mathrm{MHz}$\\
Observation duration & $T=20~\mu\mathrm{s}$\\
Energy-to-noise-density ratio & $E/N_0=0~\mathrm{dB}$\\
Scattering coefficient & $|\alpha|=1$\\
Monte Carlo trials & $Q=500$\\
\hline
\end{tabular}
\end{table}

\subsection{Waveform Construction}

For the numerical experiments, zero-centered orthogonal weighted
multitone waveforms are constructed to satisfy the waveform conditions
used in the closed-form analysis. Over the observation interval $T$,
the baseband waveform transmitted by the $m$th element is
\begin{equation}
\psi_m(t)
=\sqrt{\frac{E}{T}}
\sum_{p=1}^{P}
\sqrt{\frac{w_{m,p}}{2}}
\left(
e^{j2\pi f_{m,p}t}
+
e^{-j2\pi f_{m,p}t}
\right),
\; 0\leq t<T ,
\label{eq:num_waveform}
\end{equation}
where $m=1,\ldots,M_{\rm T}$ denotes the transmit-element index,
$p=1,\ldots,P$ indexes the positive--negative tone pairs, and
$w_{m,p}$ is the normalized spectral-energy weight assigned to the
$p$th tone pair. The factor $1/2$ equally divides this energy between
the tones at $+f_{m,p}$ and $-f_{m,p}$.

The tone frequencies are selected from the Fourier grid as
$f_{m,p}=k_{m,p}/T$, where $k_{m,p}$ is an integer frequency index and
$1/T$ is the grid spacing. Different transmit elements are assigned
disjoint sets of frequency indices to maintain waveform orthogonality.
The element-dependent FDA offsets
$\Delta f_m=i_{{\rm T},m}\Delta f$ are also aligned with the same
Fourier grid so that the frequency-shifted waveforms remain orthogonal
over $T$.

For each transmit element, the weights are chosen to satisfy
$\sum_{p=1}^{P} w_{m,p}=1$
and $\sum_{p=1}^{P} w_{m,p}f_{m,p}^{2}=\beta^{2}$
for
$w_{m,p}\geq0$,
where $\beta$ is the prescribed RMS bandwidth. The first constraint
normalizes the waveform energy, while the second fixes its spectral
second moment. Together with the symmetric tone pairs in
\eqref{eq:num_waveform}, these constraints yield
$E_m=E$, $\bar f_m=0$, and $\beta_m=\beta$ for all transmit elements.
Among the feasible weight sets, the one closest to uniform
spectral-energy allocation is used. The tone support determines the
occupied waveform bandwidth, and the weights provide the final
adjustment required to realize the prescribed RMS bandwidth.

\subsection{Validation of Modeling Approximations}
\label{subsec:assumption_validation}

Before evaluating the CRLB and estimator performance, we numerically
examine the principal approximations used in the signal model and
closed-form derivation. For the residual FDA--aperture phase neglected
in \eqref{eq:residual FDA--aperture phase}, the maximum phase magnitude
over the parameter ranges used in the subsequent numerical experiments
is $1.72\times10^{-2}$ rad ($0.985^\circ$). This small value supports
the approximation
$e^{j2\pi\Delta f_m d_{n,m}\sin\theta/c_0}\simeq 1$
throughout the considered operating range.

The waveform-orthogonality condition in
\eqref{eq:orthogonality} is evaluated using the normalized
cross-waveform metric
\begin{equation}
\epsilon_{\rm o}
\triangleq
\max_{\substack{m\neq\ell\\a,b}}
\frac{
\left|
\int
\psi_m^{(a)}(t)
\left[
\psi_\ell^{(b)}(t-\Delta\tau_{m\ell})
\right]^*
e^{j2\pi(\Delta f_m-\Delta f_\ell)t}
\,dt
\right|
}{
\left\|\psi_m^{(a)}\right\|_2
\left\|\psi_\ell^{(b)}\right\|_2
}.
\notag
\end{equation}
The projection approximations in \eqref{eq:G_def_modelII} are assessed by
\begin{equation}
\epsilon_r
\triangleq
\frac{|\mathbf s^{\rm H}\mathbf s_r|^2}
{(\mathbf s^{\rm H}\mathbf s)\|\mathbf s_r\|_2^2},
\qquad
\epsilon_\theta
\triangleq
\frac{|\mathbf s^{\rm H}\mathbf s_\theta|^2}
{(\mathbf s^{\rm H}\mathbf s)\|\mathbf s_\theta\|_2^2}.
\notag
\end{equation}
Over all evaluated parameter settings, these three indicators can be approximated as zeros, confirming that the neglected
cross-waveform and projection terms are numerically negligible.

\begin{figure}[!t]
    \centering
    \includegraphics[width=0.78\linewidth]{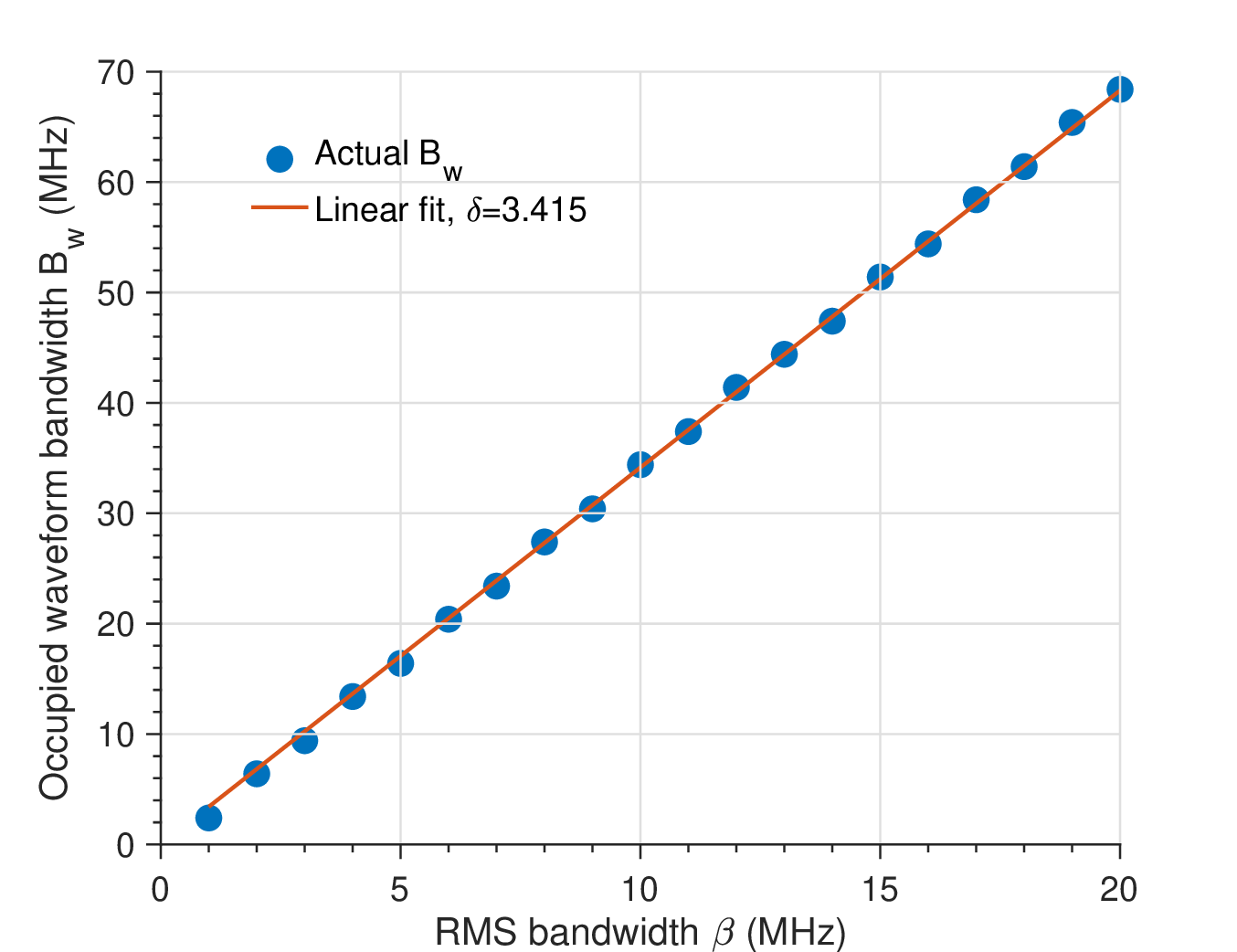}
    \caption{Occupied waveform bandwidth versus RMS bandwidth.}
    \label{fig:delta_fit}
\end{figure}

Finally, the spectral-allocation analysis in
Section~\ref{subsec:spectral_tradeoff} employs the approximation
$B_{\rm w}(\beta)\simeq\delta\beta$. Fig.~\ref{fig:delta_fit} compares
the occupied bandwidth obtained from the weighted-multitone waveform
with a zero-intercept linear fit over $\beta=1$--$20$ MHz. The fitted
coefficient is $\delta=3.415$, with a mean relative fitting error of
$3.676\%$. The close agreement between the measured bandwidths and the
linear fit supports the use of a constant $\delta$ in the analytical
spectral-allocation model.

\begin{figure}[!t]
    \centering
    \includegraphics[width=0.8\linewidth]{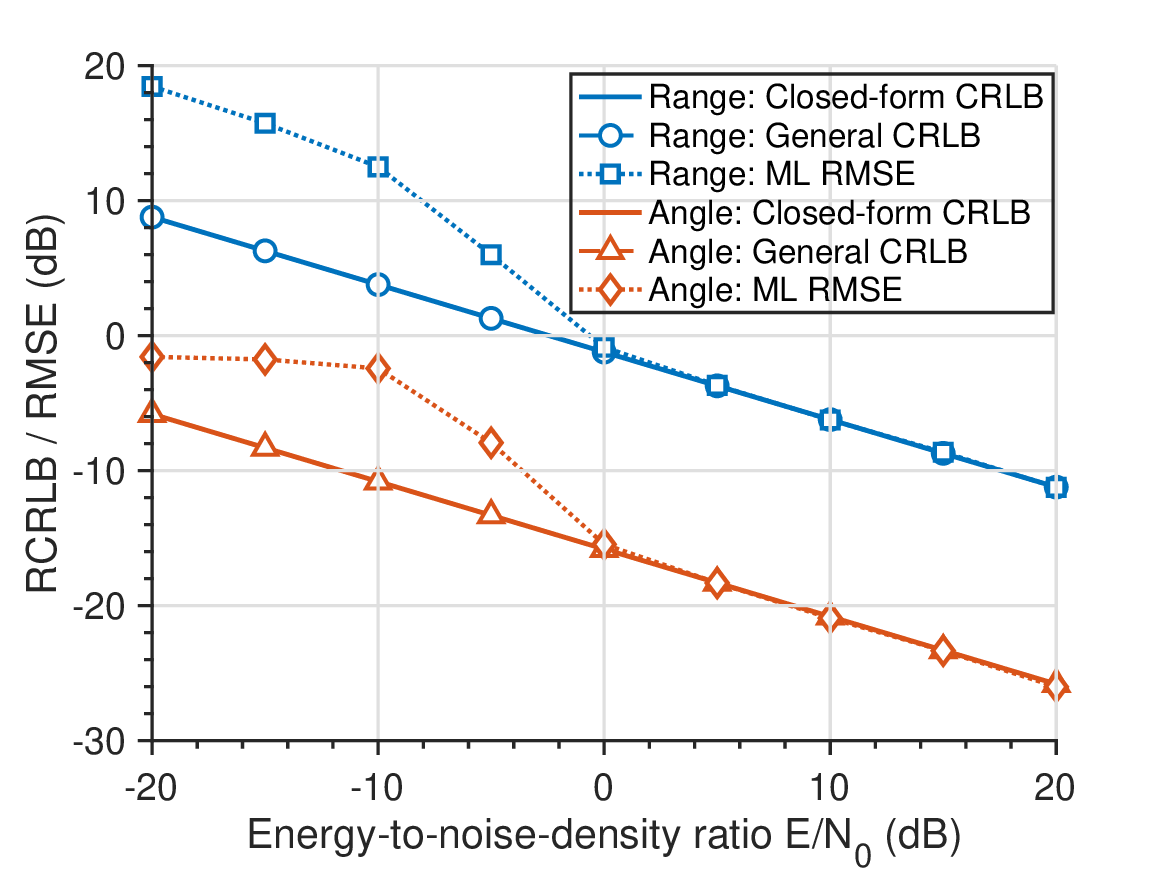}
    \caption{Closed-form and general RCRLBs and ML RMSE versus energy-to-noise-density ratio.}
    \label{fig:CRLB_MSE}
\end{figure}
\subsection{CRLB and Estimator Validation}
\label{subsec:crlb_estimator_validation}

The range and angle estimates are obtained
from the concentrated maximum-likelihood (ML) criterion
\begin{equation}
(\hat r,\hat\theta)
=
\arg\max_{r,\theta}
\frac{
\left|\mathbf{s}^{\rm H}(r,\theta)\mathbf{y}\right|^2
}{
\mathbf{s}^{\rm H}(r,\theta)\mathbf{s}(r,\theta)
}.
\label{eq:ML_concentrated}
\end{equation}
For each $\gamma_E=E/N_0$, the range--angle search region is
constructed from approximate confidence intervals based on
the corresponding closed-form RCRLBs under a Gaussian
error assumption. Each RMSE point is obtained from $Q=500$ independent realizations.

Fig.~\ref{fig:CRLB_MSE} compares the proposed unified CRLB,
closed-form CRLB, and concentrated ML estimation.
The unified and closed-form RCRLBs nearly coincide over the simulated
range, while the ML root-mean-square errors approach the corresponding
bounds at moderate and high $E/N_0$. The larger deviation at low
$E/N_0$ reflects the low-SNR threshold behavior of the ML estimator. These
results support the closed-form approximation and its use in the
following parameter studies.

\begin{figure*}[!t]
    \centering
    \subfloat[Range RCRLB]{
        \includegraphics[width=0.31\textwidth]{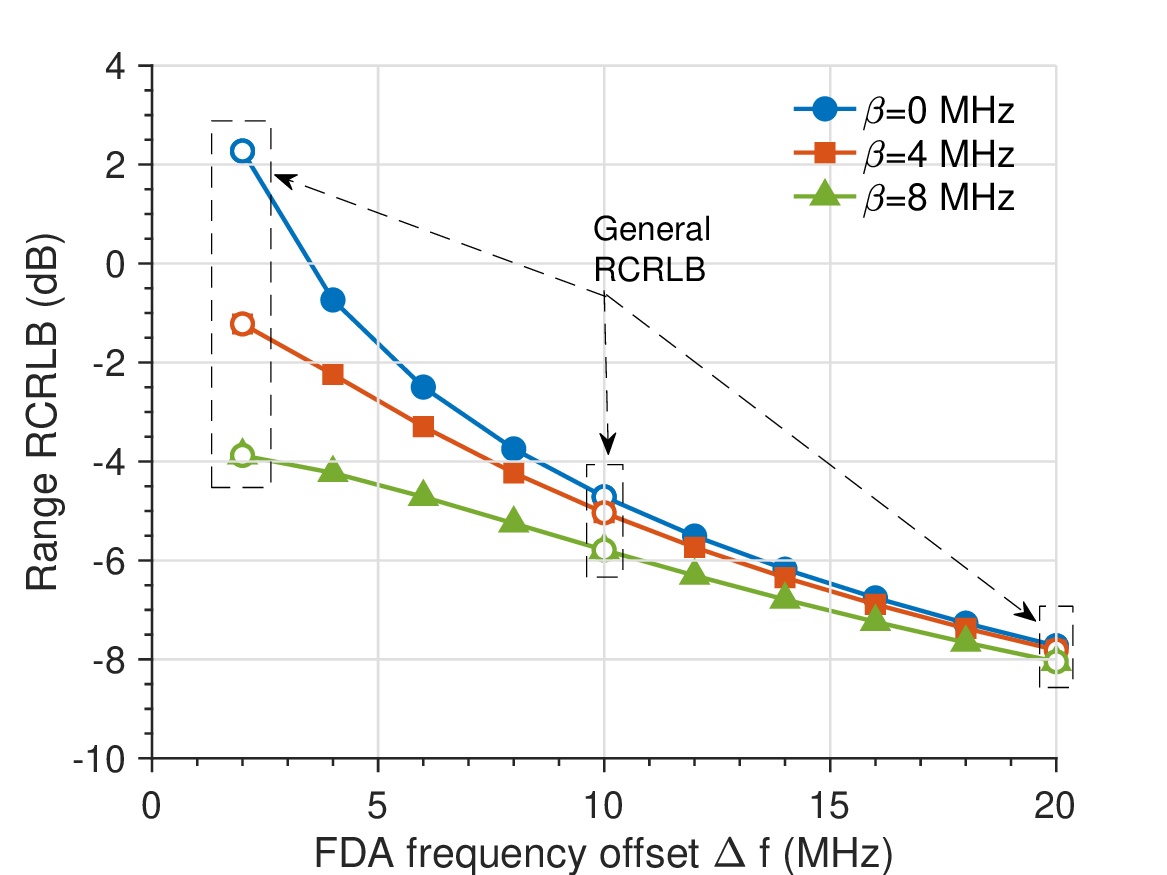}
        \label{fig:range_FrequencyOffset}
    }
    \hfill
    \subfloat[Angle RCRLB]{
        \includegraphics[width=0.31\textwidth]{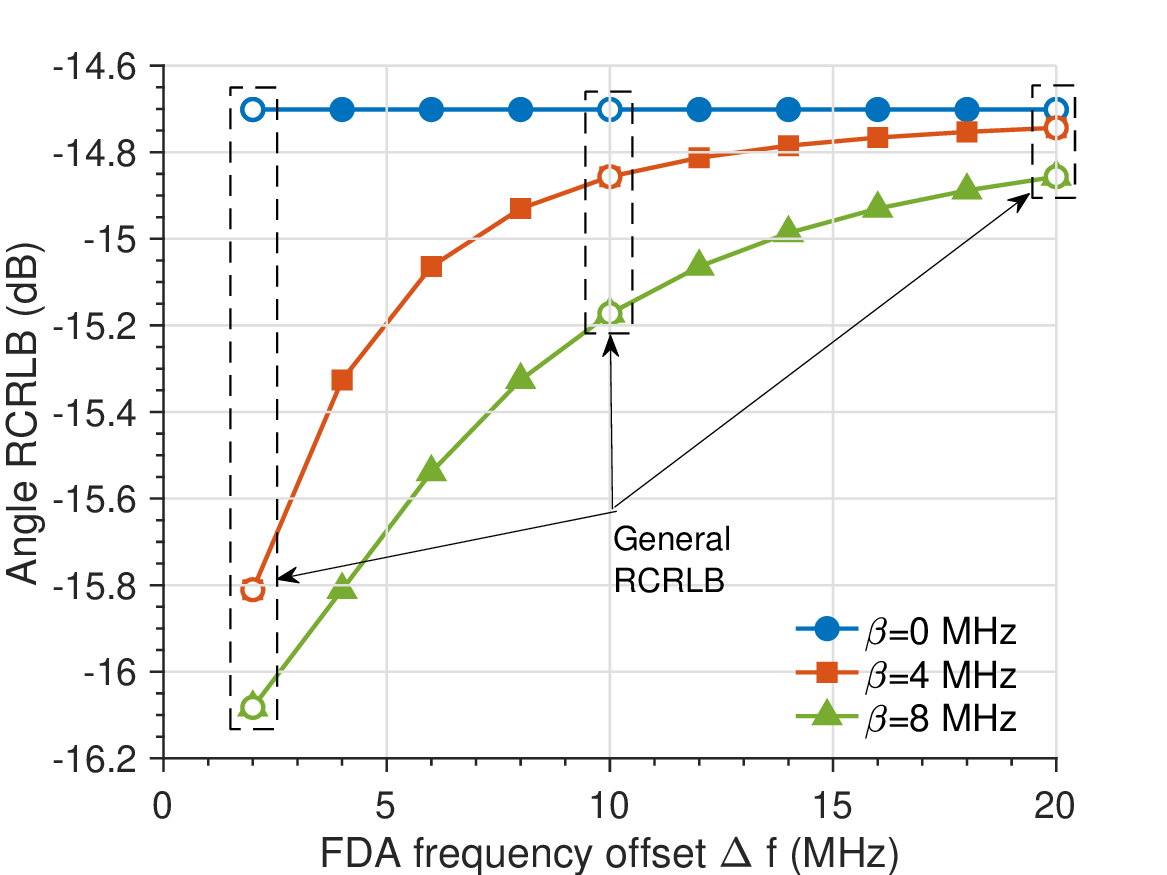}
        \label{fig:angle_FrequencyOffset}
    }
    \hfill
    \subfloat[Coupling factor]{
        \includegraphics[width=0.31\textwidth]{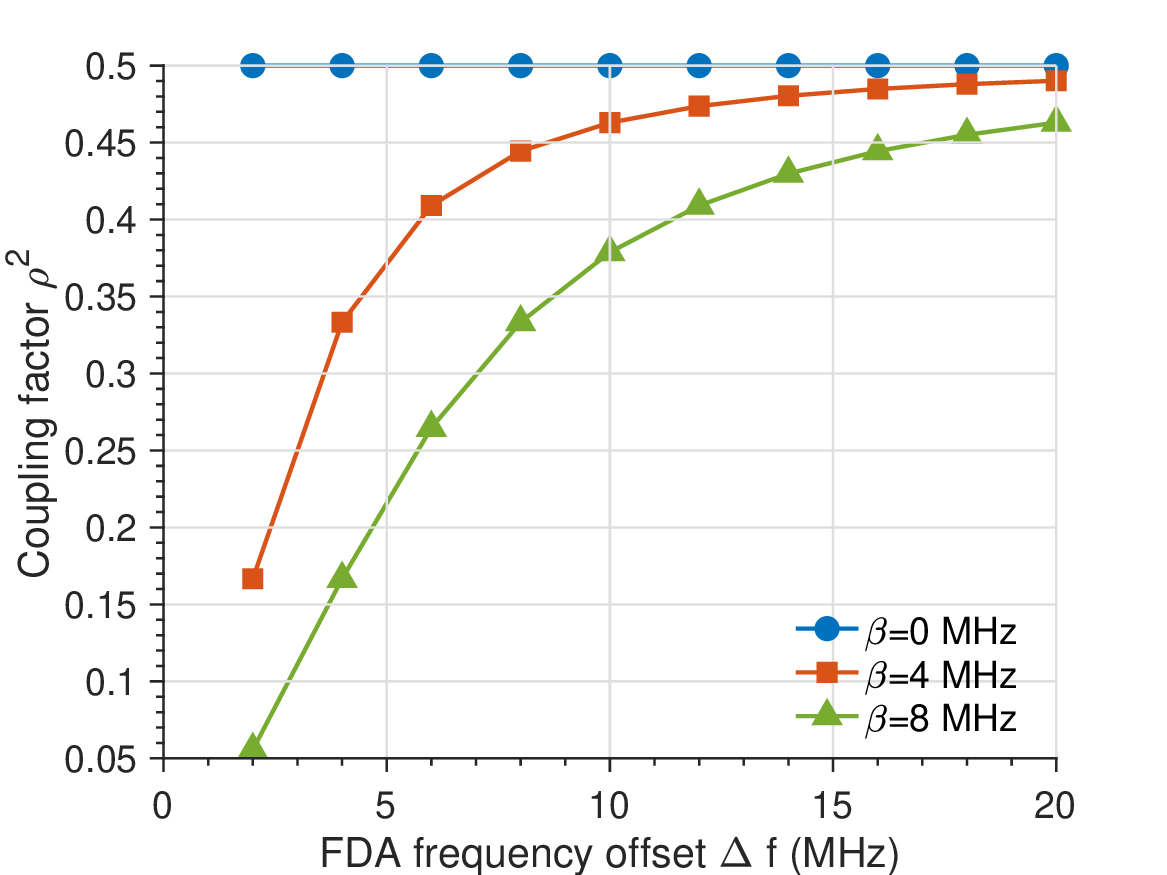}
        \label{fig:Coupling_FrequencyOffset}
    }
    \caption{Range and angle RCRLBs and range--angle coupling factor
versus FDA frequency offset.}
    \label{fig:CRLB_FrequencyOffset}
\end{figure*}

\begin{figure*}[!t]
    \centering
    \subfloat[Range RCRLB]{
        \includegraphics[width=0.31\textwidth]{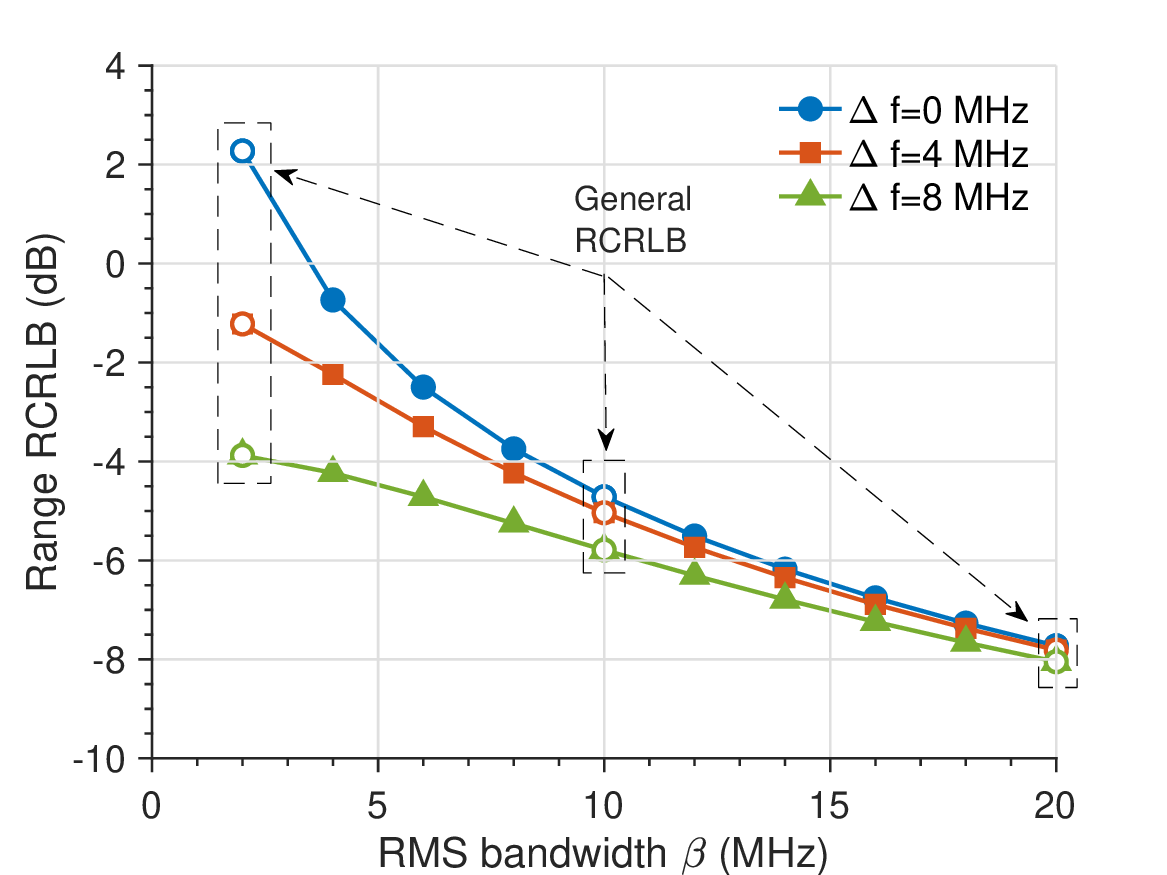}
        \label{fig:range_beta}
    }
    \hfill
    \subfloat[Angle RCRLB]{
        \includegraphics[width=0.31\textwidth]{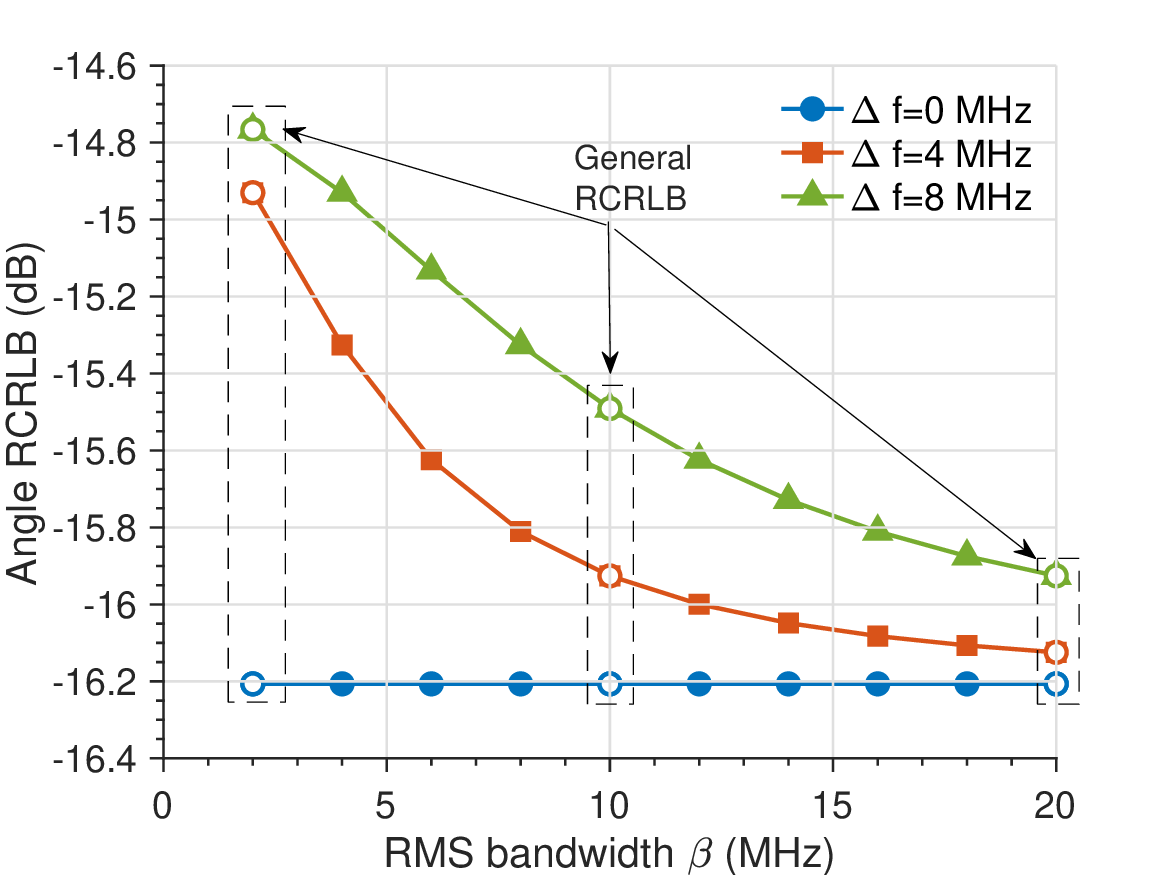}
        \label{fig:angle_beta}
    }
    \hfill
    \subfloat[Coupling factor]{
        \includegraphics[width=0.31\textwidth]{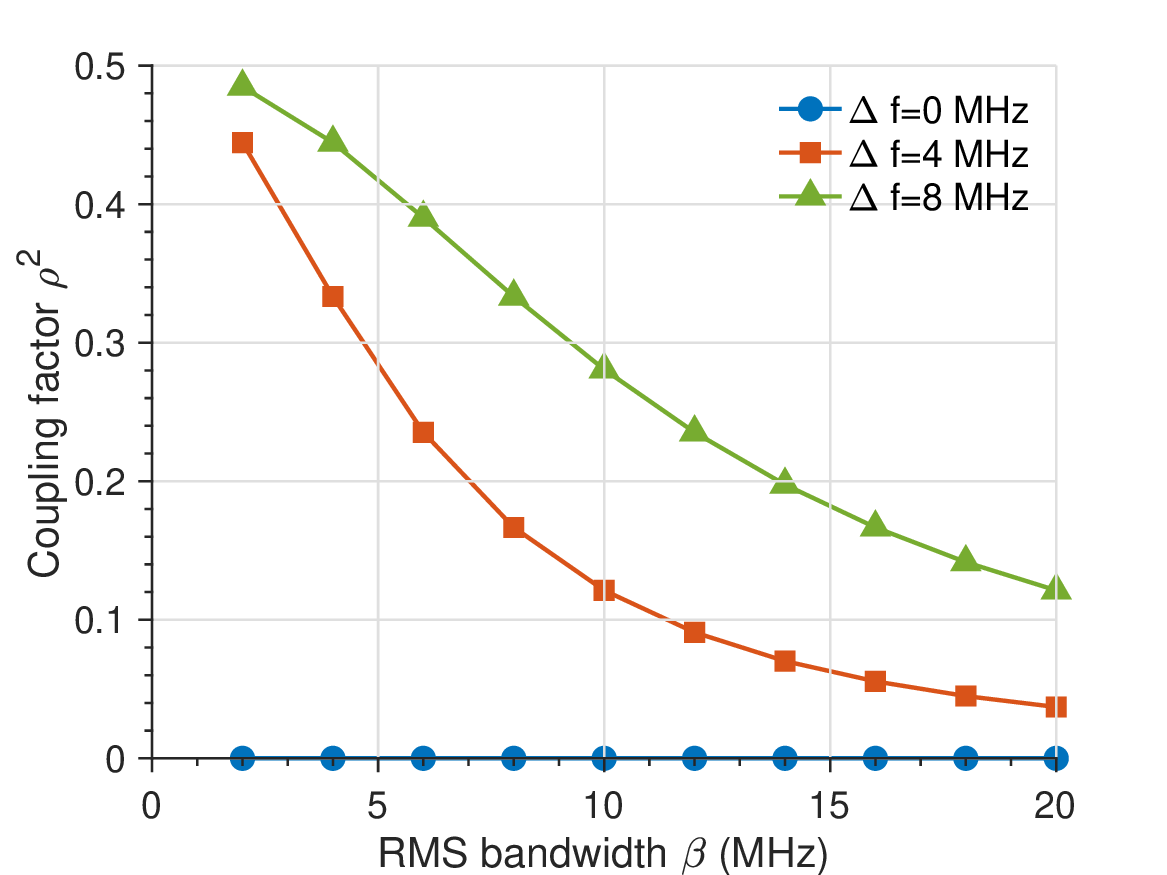}
        \label{fig:Coupling_Beta}
    }
    \caption{Range and angle RCRLBs and range--angle coupling factor
versus RMS bandwidth}
    \label{fig:CRLB_Beta}
\end{figure*}

\subsection{Bandwidth--Frequency-Diversity Effects}
\label{subsec:independent_spectral_effects}

Figs.~\ref{fig:CRLB_FrequencyOffset} and
\ref{fig:CRLB_Beta} examine the two range-information mechanisms by
varying the FDA frequency increment and waveform RMS bandwidth,
respectively. The main observations are summarized as follows. 
Note that the $\beta=0$ curve is included only as the idealized narrowband
closed-form reference \eqref{eq:NB FDA-MIMO range} and \eqref{eq:NB FDA-MIMO angle}, as introduced in \cite{Xu2015TSP}.

\emph{1) Range accuracy:}
For the frequency-offset sweep, the range RCRLB decreases monotonically
with $\Delta f$, and the selected unified-CRLB samples closely follow
the closed-form curves. Increasing $\beta$ shifts the curves downward
because the waveform-delay contribution is strengthened. The
RMS-bandwidth sweep shows the complementary behavior: increasing
$\beta$ reduces the range RCRLB for all considered frequency offsets,
while a larger $\Delta f$ provides an additional FDA-induced ranging
gain. Thus, $\beta$ and $\Delta f$ both improve range accuracy when
varied independently.

\emph{2) Angle accuracy:}
The angle RCRLB increases with $\Delta f$ when $\beta>0$ and
gradually approaches its FDA-dominant limiting behavior. In contrast,
increasing $\beta$ improves the angle RCRLB for nonzero $\Delta f$,
whereas the variation is weak when $\Delta f=0$ because the angular
information is already carrier dominated. Hence, FDA frequency
diversity and waveform RMS bandwidth have opposite effects on angular
accuracy.

\emph{3) Coupling behavior:}
Figs.~\ref{fig:Coupling_FrequencyOffset} and
\ref{fig:Coupling_Beta} show that increasing $\Delta f$ strengthens
the range--angle coupling, whereas increasing $\beta$ weakens it. This
coupling behavior is consistent with the opposite angular trends
observed in Figs.~\ref{fig:angle_FrequencyOffset} and
\ref{fig:angle_beta} under the narrowband regime with $\beta\ll f_c$.

\begin{figure*}[!t]
    \centering
    \subfloat[Range RCRLB]{
        \includegraphics[width=0.31\textwidth]{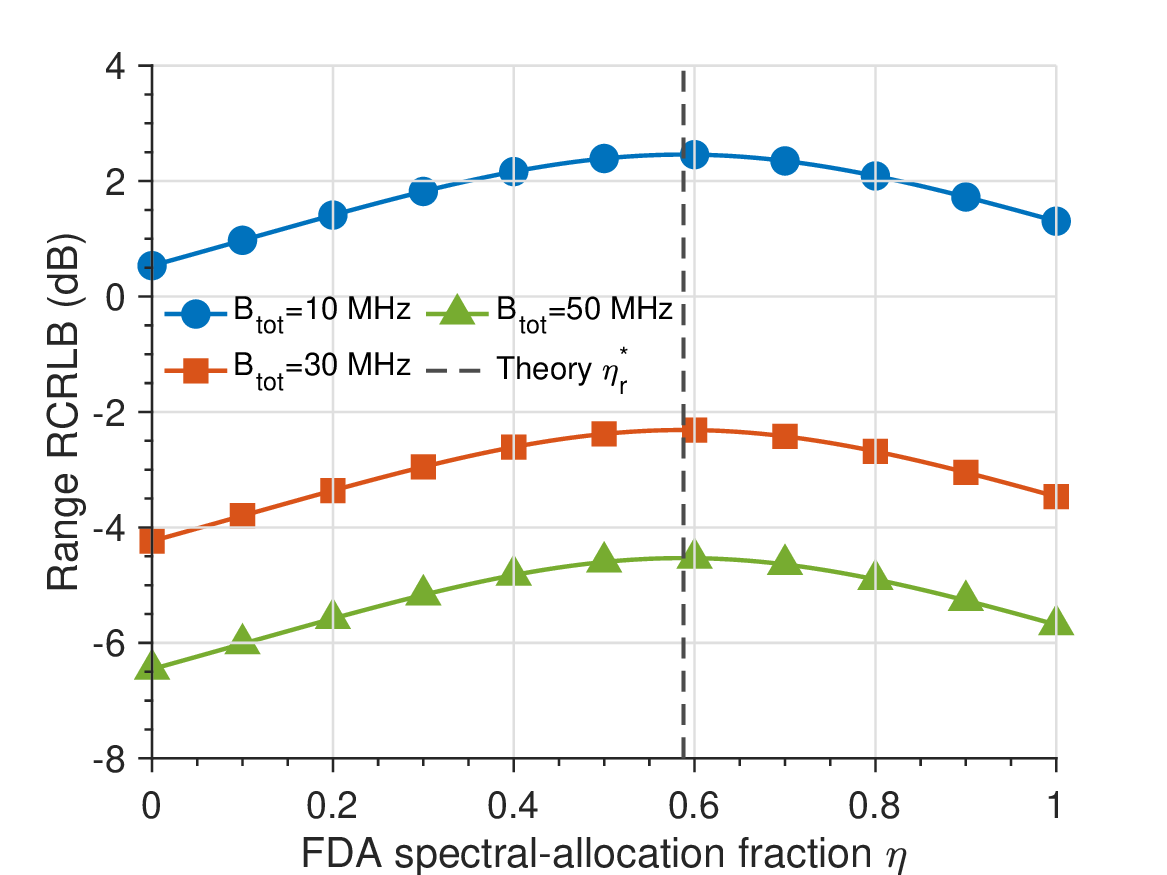}
        \label{fig:RangeCRLB_tradeoff}
    }
    \hfill
    \subfloat[Angle RCRLB]{
        \includegraphics[width=0.31\textwidth]{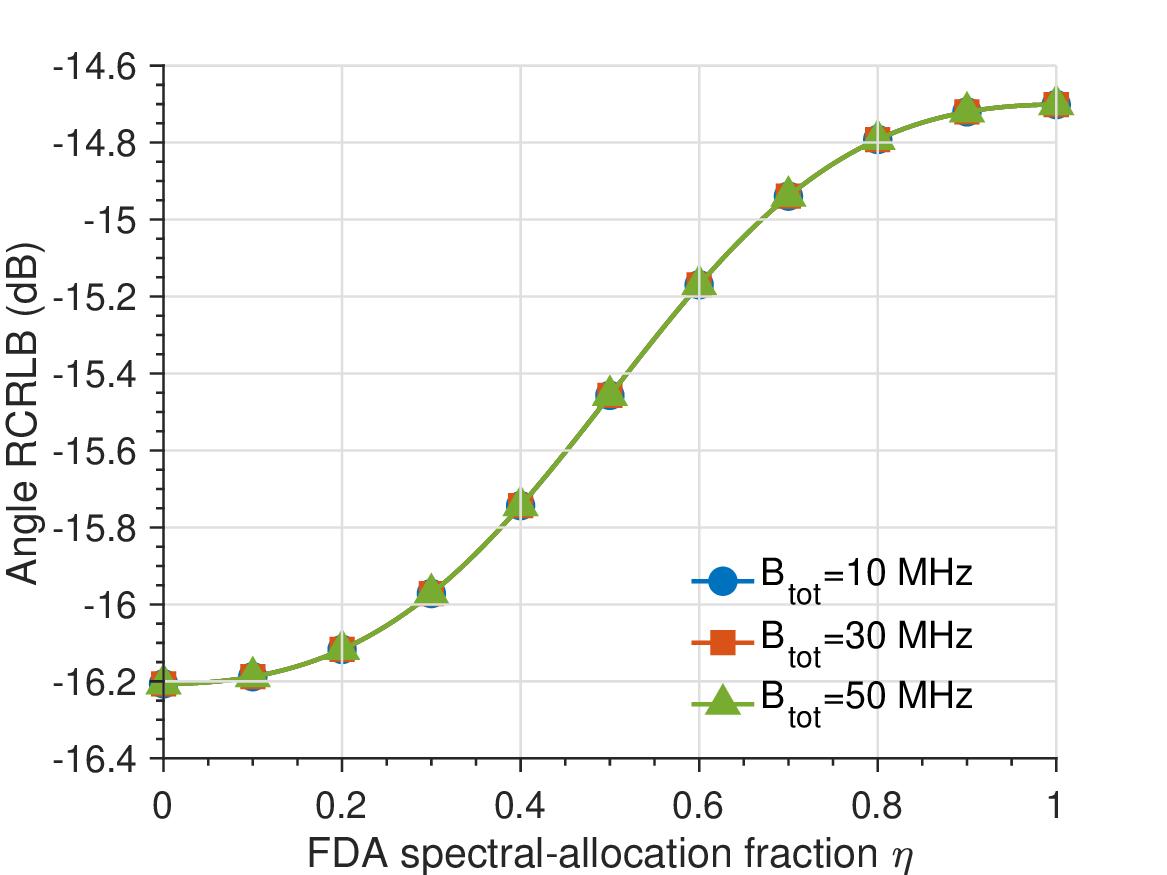}
        \label{fig:AngleCRLB_tradeoff}
    }
    \hfill
    \subfloat[Coupling factor]{
        \includegraphics[width=0.31\textwidth]{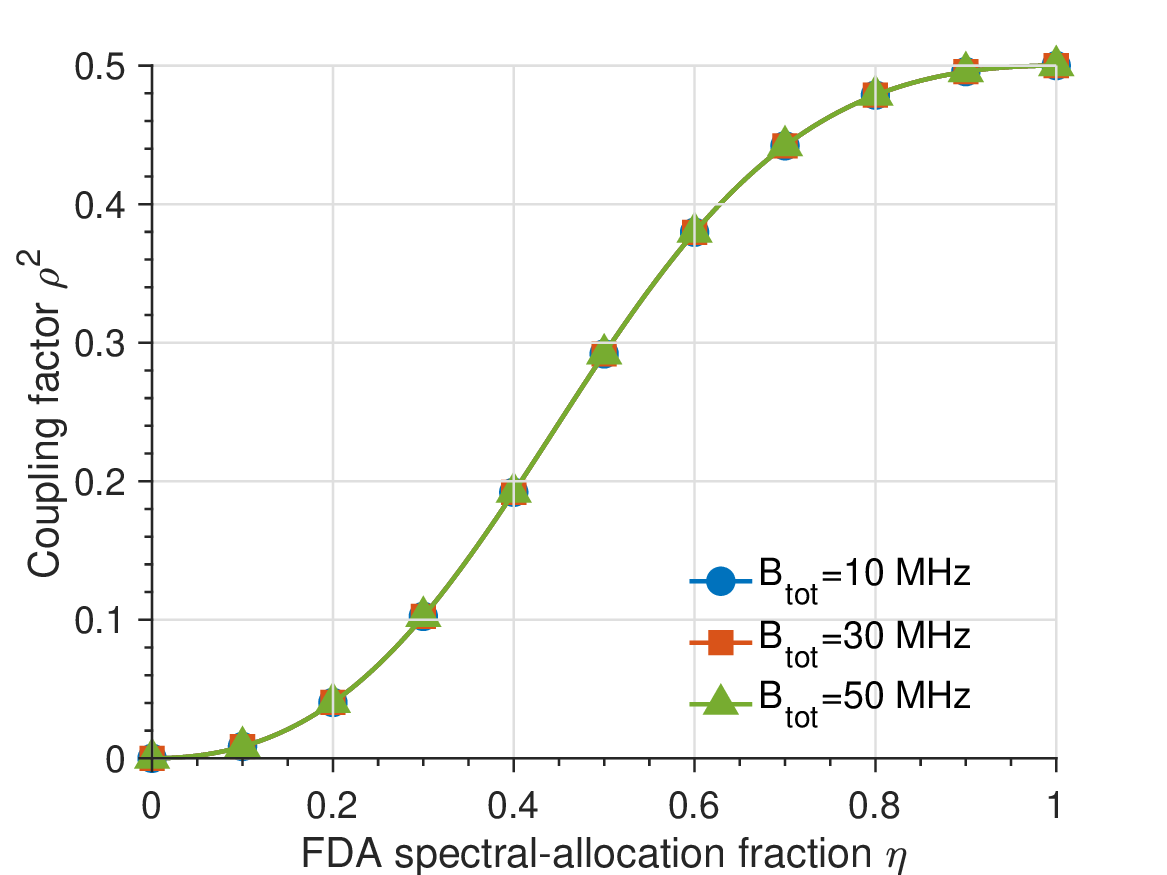}
        \label{fig:Coupling_tradeoff}
    }
    \caption{Range and angle RCRLBs and range--angle coupling factor versus the FDA
spectral-allocation fraction.}
    \label{fig:CRLB_tradeoff}
\end{figure*}

\subsection{Spectral Resource Effects}
\label{subsec:spectral_allocation_numerical}

Fig.~\ref{fig:CRLB_tradeoff} examines the allocation of a fixed
aggregate spectral budget between waveform bandwidth and FDA frequency
span.

\begin{itemize}
    \item \emph{Spectral allocation:}
    The range RCRLB in Fig.~\ref{fig:RangeCRLB_tradeoff} first
    increases and then decreases as more spectral resource is assigned
    to the FDA frequency span. The theoretical stationary point marked
    in the figure agrees well with the observed turning point, which corresponds to the maximum range RCRLB and therefore represents the worst allocation rather than an optimum. This
    behavior reflects the initial loss of waveform-delay information
    followed by the increasing contribution of FDA-induced phase
    information. The similar turning locations for different spectral budgets indicate that the stationary allocation is governed mainly by the relative partition between waveform bandwidth and FDA frequency span as shown in \eqref{eq:eta_range_stationary}, whereas the total budget primarily determines the attainable range-accuracy level.

    \item \emph{Angle and coupling costs:}
    The angle RCRLB in Fig.~\ref{fig:AngleCRLB_tradeoff} increases
    with the FDA allocation, and the coupling factor in
    Fig.~\ref{fig:Coupling_tradeoff} follows the same trend. The coupling curves for different spectral budgets nearly overlap as the normalized balance between waveform bandwidth and FDA frequency spread is governed mainly by the allocation ratio. The weak separation among the angle curves is due to the carrier-dominated angular information in the considered narrowband regime.
\end{itemize}

These results show that the competition between waveform-delay and
FDA-induced range information appears when the two mechanisms share a
finite spectral resource.

\begin{figure}[!t]
    \centering
    \includegraphics[width=1\linewidth]{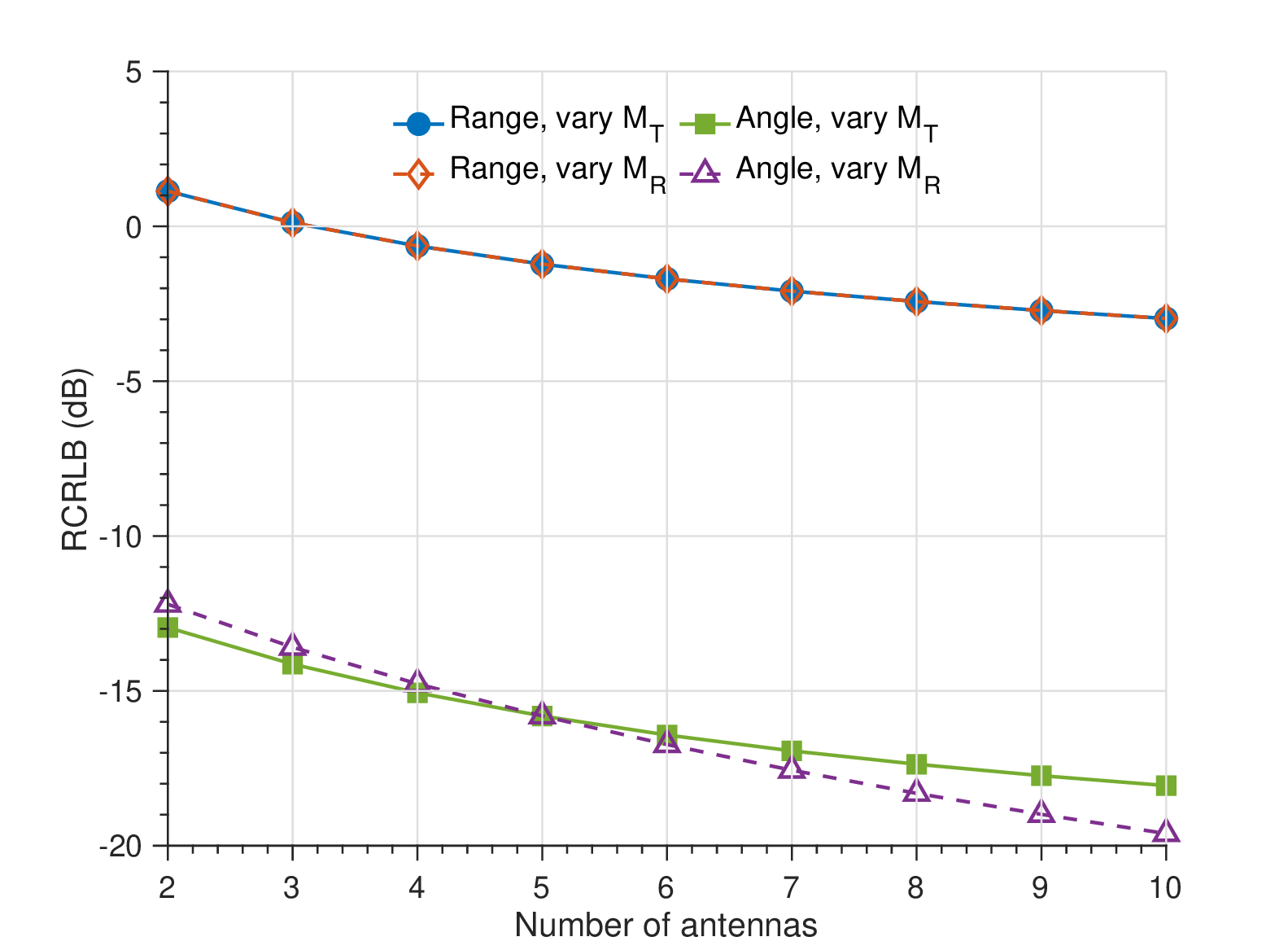}
    \caption{Range and angle RCRLBs versus transmit and receive array sizes.}
    \label{fig:CRLB_array}
\end{figure}

\begin{figure}[!t]
    \centering
    \includegraphics[width=1\linewidth]{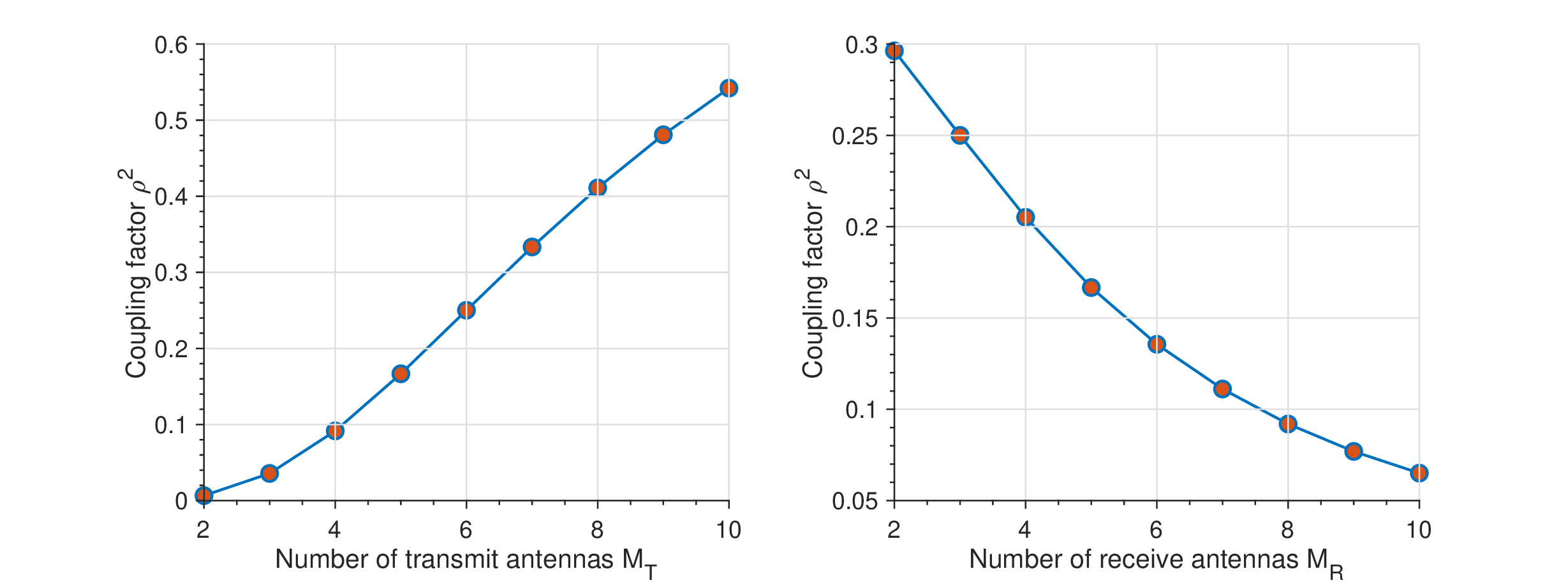}
    \caption{Coupling factor versus transmit and receive array sizes.}
    \label{fig:Coupling_ArraySize}
\end{figure}

\subsection{Array, Carrier-Frequency, and Target-Angle Effects}
\label{subsec:geometric_numerical}

The array results in Figs.~\ref{fig:CRLB_array} and
\ref{fig:Coupling_ArraySize} show that increasing either the transmit
or receive array size reduces both range and angle RCRLBs. The nearly overlapping range curves reflect the approximate symmetry
between the transmit- and receive-aperture contributions to range
information under the adopted array configuration. In contrast, the stronger angular improvement is obtained by increasing the receive
array size. This is because receive-array enlargement provides consistently favorable changes in
channel number, receive aperture, and coupling, whereas transmit-array
enlargement also increases the FDA-induced coupling. 

\begin{figure}[!t]
    \centering
    \includegraphics[width=0.7\linewidth]{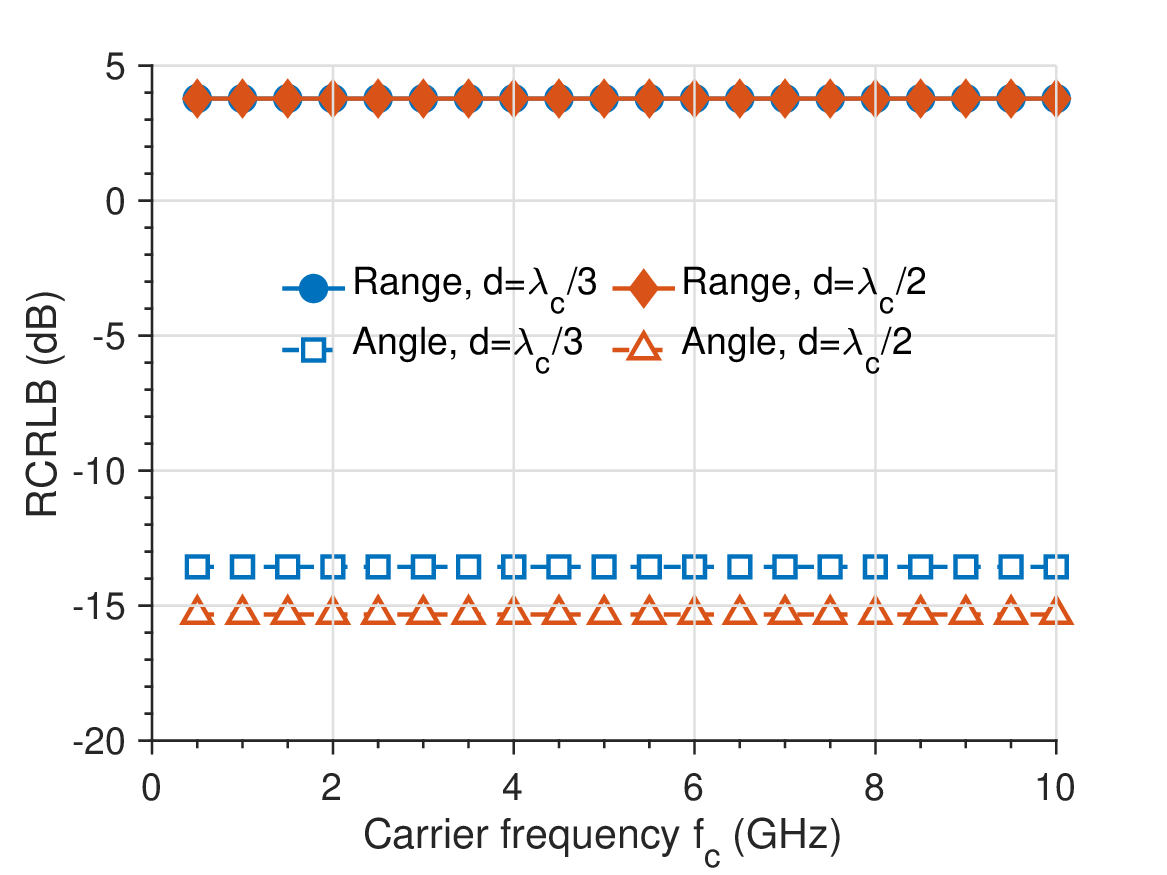}
    \caption{Range and angle RCRLBs versus reference carrier frequency.}
    \label{fig:CRLB_Fc}
\end{figure}

Fig.~\ref{fig:CRLB_Fc} shows that the range and angle RCRLBs are
nearly invariant with carrier frequency when the element spacings are
normalized to the carrier wavelength. The range curves nearly overlap as a common scaling of the
transmit and receive apertures leaves their relative geometric
contribution essentially unchanged. In contrast, the half-wavelength spacing gives a lower angle RCRLB
because of its larger electrical aperture.

\begin{figure}[!t]
    \centering
    \includegraphics[width=0.7\linewidth]{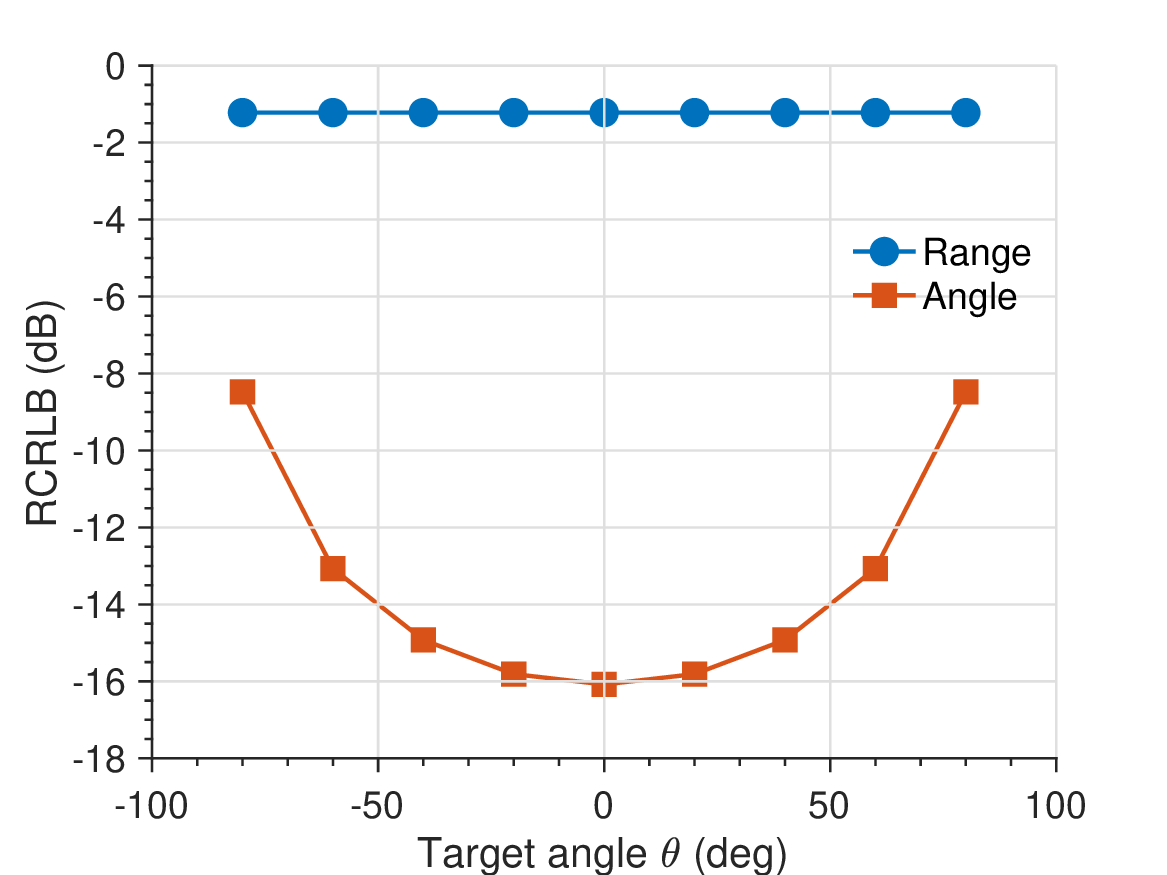}
    \caption{Range and angle RCRLBs versus target angle.}
    \label{fig:CRLB_theta}
\end{figure}

Finally, Fig.~\ref{fig:CRLB_theta} shows that the range RCRLB is
essentially independent of target angle, while the angle RCRLB is
minimized at broadside and increases symmetrically toward endfire.

Overall, the numerical results agree with the parameter trends derived
in Section~\ref{sec:CRLB_analysis}: waveform RMS bandwidth and FDA
frequency diversity both improve range accuracy when varied
independently, but they affect range--angle coupling and angular
accuracy in opposite directions. Their competition becomes explicit
when a finite aggregate spectral budget is imposed.

\section{Conclusion}
\label{sec:conclusion}

In this paper, we developed a bandwidth-aware CRLB framework for joint
range--angle estimation in colocated monostatic FDA-MIMO radar. The
sampled-signal formulation retains waveform-delay sensitivity and
FDA-induced phase diversity, while the effect of the unknown
quasi-static complex scattering coefficient is accounted for through
an EFIM. Under the stated waveform and array conditions and a centered
linear frequency-offset law, the resulting closed-form CRLBs explicitly
characterize the interplay between waveform RMS bandwidth and FDA
frequency diversity. A three-factor decomposition of the squared
normalized cross-information further explains their contributions to
range--angle coupling.

When increased independently, either waveform RMS bandwidth or FDA
frequency diversity reduces the range CRLB. Their angular effects
differ: increasing waveform RMS bandwidth reduces the angle CRLB,
whereas increasing FDA frequency diversity increases it for nonzero
waveform RMS bandwidth. Under a fixed aggregate spectral budget, the range CRLB
reaches an interior maximum and attains its minimum at an allocation
boundary, whereas the angle CRLB increases monotonically with the
FDA spectral share. Numerical comparisons of the general and
closed-form bounds, together with maximum-likelihood estimation results,
support these predictions under the considered conditions. Future work
may address nonuniform frequency offsets, joint range--angle--Doppler
estimation, and robustness to practical RF impairments.

\appendices
\normalsize

\section{Conditional-Average FIM Interpretation for Sample-Varying RCS}
\label{sec:fast_extension}

The main derivation assumes a scattering coefficient that is
constant over one coherent processing interval. To examine whether the
same parameter dependencies appear under rapid fluctuations, we
consider a representative sample-varying scattering model and average
the FIM conditioned on the scattering realization. 

Let the complex scattering coefficient vary across samples as \cite{3gpp_R1_2500414}
\begin{equation}
\alpha_k = \sqrt{A}\sqrt{B_k}e^{j\phi_k},
\end{equation}
where $A$ denotes the large-scale mean RCS, $B_k$ is a unit-mean small-scale fluctuation. We assume that $\phi_k$ is independently and uniformly distributed
over $[0,2\pi)$ and is independent of $B_k$, with
$\mathbb E[B_k]=1$. The sampled received signal can be written as
\begin{equation}
\mathbf{y}=\mathbf{D}(\boldsymbol{a})\mathbf{s}+\mathbf{w},
\end{equation}
where $\boldsymbol{a}=[\alpha_0,\ldots,\alpha_{K-1}]^T$ and
\[
\mathbf{D}(\boldsymbol{a})
=
\mathbf{I}_{M_{\rm R}}\otimes \mathrm{diag}(\alpha_0,\ldots,\alpha_{K-1}),
\]
where $\otimes$ is the Kronecker product operator.
Assume $\mathbb{E}[\alpha_k]=0$ and
$\mathbb{E}[\alpha_k\alpha_\ell^*]=A\delta_{k\ell}$ with $\delta_{kk}=1$ and $\delta_{k\ell}=0$ for $k\neq\ell$. Since the scattering process is independent of
$\mathbf{p}=[r,\theta]^T$, the expected conditional FIM can be expressed as
\begin{equation}
[\mathbf{J}_{\rm c}(\mathbf{p})]_{ij}
=
\mathbb{E}_{\boldsymbol{a}}
\left[
\frac{2}{\sigma^2}
\Re\left\{
\left(\frac{\partial \boldsymbol{\mu}}{\partial p_i}\right)^H
\left(\frac{\partial \boldsymbol{\mu}}{\partial p_j}\right)
\right\}
\right],
\end{equation}
with $\boldsymbol{\mu}=\mathbf{D}(\boldsymbol{a})\mathbf{s}$. Using the second-order moment of $\alpha_k$ gives
\begin{equation}
\mathbf{J}_{\rm c}(\mathbf{p}) = A\mathbf{G}_{0}(r,\theta),
\end{equation}
where $\mathbf{G}_{0}(r,\theta)$ has the same structural form as the simplified FIM in Section~\ref{sec:CF CRLB}. Therefore,
\begin{equation}
\mathrm{CRLB}_{\rm c}(\mathbf{p})
\approx
\frac{1}{A}\mathbf{G}_{0}^{-1}(r,\theta).
\end{equation}

Thus, averaging the conditional information changes the
scattering-power factor from $|\alpha|^2$ to $A$ but preserves the
leading-order dependence on waveform RMS bandwidth, FDA frequency
offset, carrier frequency, and array geometry. 

\bibliographystyle{IEEEtran}
\bibliography{MyFDA_CRLB1}

\end{document}